\documentclass[11pt,a4paper]{article}
\usepackage[margin=1.5in]{geometry}
\pdfoutput=1
\usepackage{amsthm,amsmath,amssymb}
\usepackage{authblk}

\usepackage[authoryear]{natbib}
\usepackage[bookmarks=false]{hyperref}
\usepackage[utf8]{inputenc} 

\usepackage{graphicx}

\usepackage{subfigure}
\usepackage{url}
\newcommand{\btheta}{\boldsymbol \theta}
\newcommand{\bR}{\mathbf R}
\newcommand{\br}{\mathbf r}
\newcommand{\bZ}{\mathbf Z}
\newcommand{\bT}{\mathbf T}
\newcommand{\bz}{\mathbf z}
\newcommand{\hao}{\hat \alpha^o}
\newcommand{\amo}{\alpha^o_m}
\renewcommand{\th}{^\text{th}}
\newcommand{\prd}[1]{\prod_{#1=1}^{\MakeUppercase #1}}
\newcommand{\sumover}[1]{\sum_{#1=1}^{\MakeUppercase #1}}
\newcommand{\znm}{z_{nm}}
\newcommand{\given}{\,|\,}
\newcommand{\E}[2]{\mathbb E_{#1} \Big[#2\Big]}
\renewcommand{\d}{\,\text d}
\usepackage{ulem}

\title{Fast and accurate approximate inference of transcript expression from RNA-seq data}

\author[1]{James Hensman\thanks{james.hensman@sheffield.ac.uk}$^\ddagger$}
\author[2]{Panagiotis Papastamoulis\thanks{panagiotis.papastamoulis@manchester.ac.uk}\footnote{These two authors contributed equally.}}

\author[3]{Peter Glaus}
\author[4]{Antti Honkela}
\author[2]{Magnus Rattray}
\affil[1]{Sheffield Institute for Translational Neuroscience (SITraN), UK}
\affil[2]{Faculty of Life Science, The University of Manchester, UK}
\affil[3]{School of Computer Science, The University of Manchester, UK}
\affil[4]{Helsinki Institute for Information Technology HIIT, Department of Computer Science, University of Helsinki, Finland}

\begin{document}
  \maketitle
\newcommand{\james}{\textcolor{blue}}
\newcommand{\antti}{\textcolor{blue}}


\begin{abstract} 
\noindent
\textbf{Motivation:} Assigning RNA-seq reads to their transcript of origin is a fundamental task in
transcript expression estimation. Where ambiguities in assignments exist due to
transcripts sharing sequence, e.g. alternative isoforms or alleles, the problem
can be solved through probabilistic inference. Bayesian methods have been shown
to provide accurate transcript abundance estimates compared to competing
methods. However, exact Bayesian inference  is intractable and approximate
methods such as Markov chain Monte Carlo (MCMC) and Variational Bayes (VB) are
typically used. While providing a high degree of accuracy and modelling
flexibility, standard implementations can be prohibitively slow for large
datasets and complex transcriptome annotations.\\
\textbf{Results:} We propose a novel approximate inference scheme based on VB and apply it to an existing model of transcript expression inference from RNA-seq data. Recent advances in VB algorithmics are used to improve the convergence of the
algorithm beyond the standard Variational Bayes Expectation Maximisation (VBEM)
algorithm. We apply our algorithm to simulated and biological datasets,
demonstrating a significant increase in speed with only very small loss
in accuracy of expression level estimation. We carry out a comparative study
against seven popular alternative methods and demonstrate that our new algorithm
provides excellent accuracy and inter-replicate consistency while 
remaining competitive in computation time.\\
\textbf{Availability:} The methods were implemented in R and C++, and are available as part of the
BitSeq project at \url{github.com/BitSeq}.\\ The method is also available
through the BitSeq Bioconductor package. The source code to reproduce all simulation results can be accessed via \url{github.com/BitSeq/BitSeqVB\_benchmarking}
\end{abstract}

\section{Introduction}

RNA-seq is a technology with the potential to identify and quantify
all mRNA transcripts in a biological sample~\citep{mortazavi2008}. Some of these transcripts
come from different isoforms or alleles of the same genes or
from closely related homologous genes, and
consequently they may share much of their primary sequence. Currently popular
RNA-seq technologies generate short reads that must be aligned
to the genome or transcriptome in order to quantify expression
levels. In some cases the
observed reads could originate from several different transcripts and there may be
few reads that are useful to distinguish these transcripts. It is therefore a 
challenging statistical problem to uncover the expression
levels of closely related transcripts.  A recent
assessment confirms this by showing significant
variability between results obtained using different computational 
pipelines~\citep{SEQC2014}.

Probabilistic latent variable models, in particular mixture models \citep{Jiang2009,Li2010,Katz2010,Li2011,Turro2011,glaus2012,Trapnell2013,nariai2013tigar}
provide a popular and effective approach for inferring transcript expression 
levels from RNA-seq data. Such
models can be used to deconvolve the signal in the read data, assigning reads to
alternative, pre-defined transcripts according to their probability of originating from
each. The term mixture model derives from the interpretation of the data as being
derived from a mixture of different transcripts, the mixture components, with each read
originating from one component. Although reads originate from only one
component they may map to multiple related components, resulting in some
ambiguity in their assignment. Transcript expression levels are model parameters (mixture component proportions) that have to be inferred from the mapped read data. Due to their probabilistic nature these models can fully account for multiple mapping reads, complex
biases in the sequence data, sequencing errors, alignment quality scores and
prior information on the insert length in paired-end reads. Mixture models have been
successfully applied to infer the proportion of different gene isoforms or
allelic variants in a particular sample \citep{Jiang2009,Katz2010,Turro2011}, for inferring gene and isoform expression levels \citep{Li2010,Li2011,roberts2013streaming,Trapnell2013, mortazavi2008} and for transcript-level differential expression calling~\citep{glaus2012,Trapnell2013}.

Inference in latent variable models such as these can be carried out by Maximum Likelihood (ML) or
Bayesian parameter estimation. In ML the choice of parameters that maximises
the data likelihood is obtained through a numerical optimisation procedure. In the
case of mixture models a popular choice of algorithm is the Expectation
Maximisation (EM) algorithm, as first applied to this model and
expressed sequence tag data by \cite{Xing2006} and later to RNA-seq data by \citet{Li2010}. For Bayesian
inference the most popular approach is Markov chain Monte Carlo (MCMC) and
for the case of mixture models a Gibbs sampler is most often used \citep{Katz2010,Li2011, glaus2012}. An advantage of Bayesian inference is that one obtains a posterior probability over the model
parameters rather than just a point estimate. This provides a level of uncertainty
in the inferred transcript expression levels as well as information about the covariation
between estimates for closely related transcripts. The uncertainty information can
be usefully propagated into downstream analysis of the data, e.g. calling differentially
expressed transcripts from replicated experiments~\citep{glaus2012}.

A Bayesian method, BitSeq, was proposed in which inference was carried out
using a collapsed Gibbs sampler \citep{glaus2012}. The method was shown to perform well,
especially for the task of inferring the relative expression of different gene isoforms and
for ranking transcripts according to their probability of being differentially
expressed between conditions. However, for typical modern RNA-seq datasets with
hundreds of millions of read-pairs the Gibbs sampler can be inconveniently slow, creating a
computational bottleneck in applying a Bayesian approach.
As the volume of data continues to grow and gene models are becoming more complex as more alternative transcripts are discovered, more efficient inference algorithms are required so that Bayesian methods can be used to provide practical computational tools.

An alternative approach to Bayesian inference is to use deterministic approximate inference
algorithms such as Variational Bayes (VB)~\citep[reviewed in][]{bishop2006pattern}. While MCMC algorithms are attractive due to
their asymptotic approximation guarantees, VB often provides a much faster method to obtain a
good approximation to the posterior distribution. For models where Gibbs sampling
can be applied there is typically a closely related VB Expectation Maximisation (VBEM) algorithm. In this contribution we
show how VB can be used to massively speed up inference in the BitSeq model for
transcript expression-level inference. We show that the mean transcript expression level estimates are very 
close to those obtained with MCMC. We use
a recent formulation of VB \citep{hensman2012fast} which is shown to provide a greater speed up when
compared to a more standard VBEM algorithm. Our new algorithm is implemented in the
most recent version of the BitSeq, allowing the method to be applied to much
larger RNA-seq datasets in equal computing time.

An alternative VB method, TIGAR, was recently proposed for the same 
problem using a standard VBEM
algorithm \citep{nariai2013tigar}.
The assumptions made in our approximation are similar to those used in TIGAR, but the empirical
comparisons herein show that our proposed method performs better in terms of
computation time and required memory, while also providing improved accuracy on real and simulated data. The improvement in terms of reduced computational cost is due to our adoption of a novel VB method. Furthermore we investigate the effects of the variational assumption in this problem, and compare empirically to results using the gold standard, MCMC.

The paper is organised as follows. In Section~\ref{sec:methods} we review the original BitSeq probabilistic model and describe our new inference algorithm, BitSeqVB, explaining the principles underlying our improved optimisation scheme. In Section~\ref{sec:results} we benchmark our new method against the original BitSeq algorithm and six popular alternative methods using realistic simulated data and real human RNA-Seq data. We consider accuracy in terms of expression estimation, relative with-gene transcript proportions and between-replicate consistency. We also compare the computation time required for all methods and compare the new VB algorithm to more standard MCMC and VBEM inference algorithms. 

\section{Methods}

\begin{figure}[t]
\centering
\includegraphics[width=0.4\textwidth]{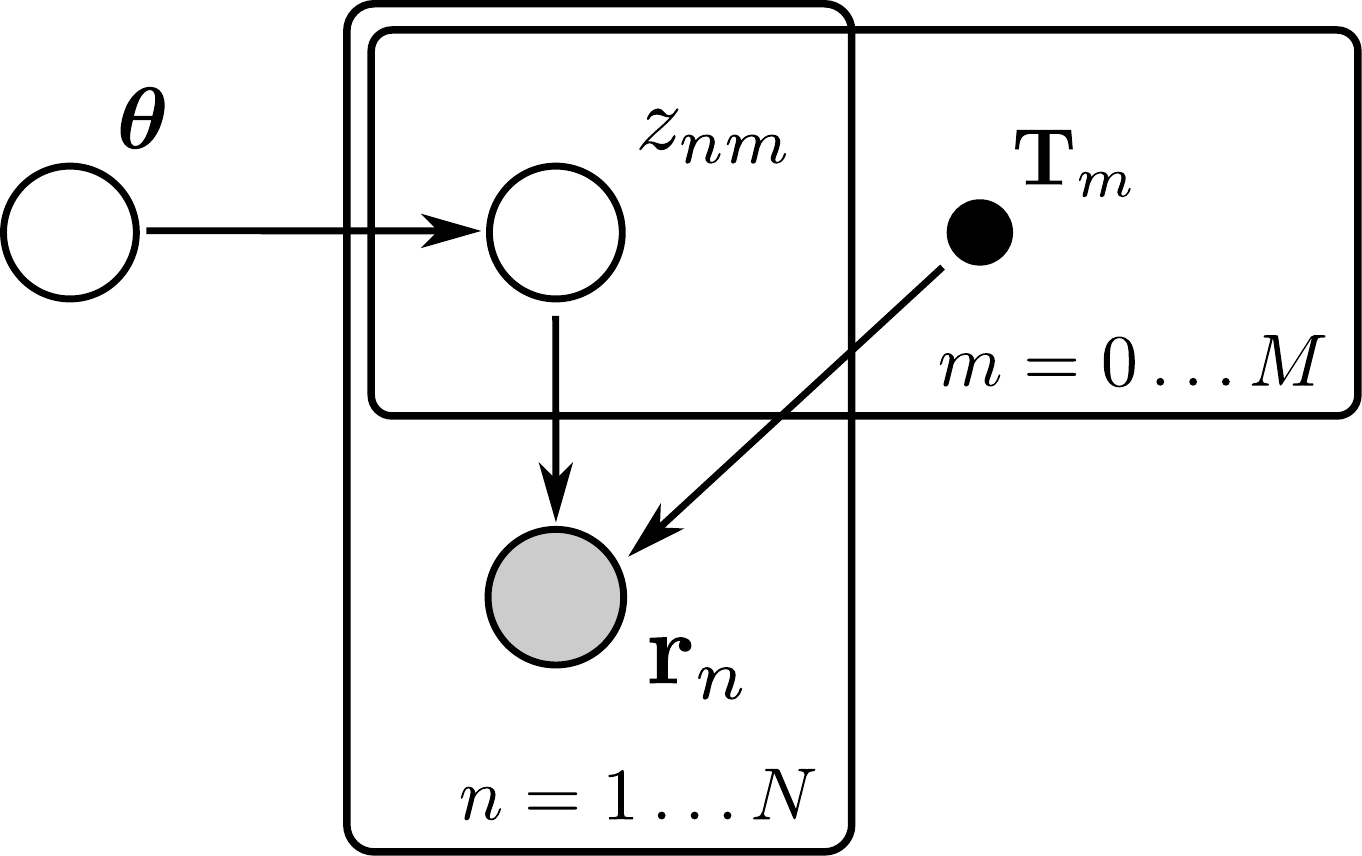}
\vspace{1.5em}
\caption{Graphical model of the RNA-seq mixture problem. Given a known Transcriptome $\bT$ and some observed reads $\bR$, the inference problem is for $\btheta$ through the latent variables $\bZ$.
}
\label{fig:graphical}
\end{figure}

Our probabilistic model of RNA-seq follows Stage 1 of \citet{glaus2012}, and is similar to that used by RSEM.
We summarise our notation in Table \ref{tab:notation}. The probabilistic model is shown using standard directed graphical notation in Figure~\ref{fig:graphical}. Here we have focussed on the mixture part of the analysis, assuming that the model which associates reads to transcripts (i.e. $p(\br_n\given \bT_m)$) is known. Following BitSeq \citep{glaus2012}, we compute this part of the model { a priori}, with parameters estimated from uniquely aligned reads.
We consider RNA-seq assays independently, computing an approximate posterior for the transcript proportions $\btheta$ in each assay.
Subsequent analysis such as differential expression can be done using the
estimated distributions of each assay.

\begin{table}[h!]
\caption{Summary of notation}\label{tab:notation}
{\begin{tabular}{cl}
$N$ & Number of reads in the dataset. \\
$M$ & Number of transcripts in the transcriptome. \\
$\br_n$ & The $n\th$ read.\\
$\bR$ & The collection of reads.\\
$\bT$ & The transcriptome. \\
$T_m$ & The $m\th$ transcript. \\
$\theta_m$ & Proportion of transcript $T_m$ in the sample. \\
$z_{nm}$ & Binary: $z_{nm}=1$ if read $n$ comes from transcript $m$. \\
$\bz_{n}$ & Allocation vector of the $n\th$ read. \\
$\bZ$ & Collection of all allocation vectors.\\
$\phi_{nm}$ & Approximate posterior probability of $z_{nm}=1$. \\
$\gamma_{nm}$ & Re-parameterisation of $\phi_{nm}$. \\
\end{tabular}}{
}
\end{table}

\subsection{The generative model}
\paragraph{Transcript fragment proportions\\}
The generative model for an RNA-seq assay is as follows. We assume that the experiment produces of collection of RNA fragments, where the abundance of fragments derived from transcript $T_m$ in the assay is $\theta_m$. Fragments are then sequenced in these proportions, so that the prior probability of any fragment corresponding to transcript $T_m$ is $\theta_m$. Introducing a convenient allocation vector $\bz_n$ for each read, we can write
\begin{equation}
p(\bZ|\btheta) = \prd n \prd m \theta_m^{\znm},
\end{equation}
where $\znm \in \{0,1\}$ is a binary variable which indicates whether the $n\th$
fragment came from the $m\th$ transcript ($\znm=1$) and is subject to $\sum_{m=0}^M \znm = 1$.
We use $\bZ$ to represent the collection of all allocation vectors.
We note that both $\btheta$ and $\bZ$ are variables to be inferred, with
$\btheta$ the main object of interest. $\btheta$ can be transformed later into
some more convenient measure, for instance reads per kilobase of length per
million sequenced reads (RPKM)~\citep{mortazavi2008}, though it is more
convenient from a probabilistic point of view to work with $\btheta$ directly.
The variables $\bZ$ are sometimes known in the machine learning literature as latent variables. Although not of interest directly, inference of these variables is essential in order to infer $\btheta$.

\paragraph{Read model\\}
An important part of the model is the likelihood term $p(\br_n|T_m)$ which is the probability of generating the $n\th$ read from the $m\th$ transcript.
Writing the collection of all reads as $\bR = \{\br_n\}_{n=1}^N$, the likelihood given a set of alignments $\bZ$ is
\begin{equation}
p(\bR|\bT,\bZ) = \prd n \prd m p(\br_n|T_m)^{\znm},
\end{equation}
where $T_m$ represents the $m\th$ transcript and $\bT$ represents the transcriptome.
The values of $p(\br_n|T_m)$ for all alignments can be computed before
performing inference in $\btheta$ since we are assuming a known transcriptome.
For paired-end reads, the mates originate from a single fragment and their
likelihood is inferred jointly. Denoting $\br_n=(r_n^{(1)},r_n^{(2)})$, the likelihood of alignment is computed as
\begin{equation}\label{eq:alignmentProbability}
P(\br_n|T_m) = P(l|T_m)P(p|l,T_m)\prod_{i=1,2}P(r_n^{(i)}|seq_{mlp}) \ ,
\end{equation}
where $l$ is the length of a fragment, $p$ is its position and $seq_{mlp}$
denotes the underlying reference sequence.
The fragment length distribution can be pre-defined or inferred empirically.
The position likelihood, $P(p|l,T_m)$, can be either uniform or
account for different biases using an empirical model as in \cite{glaus2012}.
The last term, $\prod_{i=1,2}P(r_n^{(i)}|seq_{mlp})$ describes the probability
of observed read
sequences based on quality scores and base discrepancy between read
and reference.
For detailed description of the alignment likelihood estimation please refer to
\cite{glaus2012}.

\paragraph{Identifying noisy reads\\}
Our model is similar to previous work \citep{glaus2012}, but does not contain a
variable identifying reads as belonging to a `noise' class.
To circumvent the explicit formulation of a model with this variable, we
introduce a `noise transcript' which we append to the list of known transcripts.
The generative probability of any read from this transcript, $p(\br_n|T_0)$, is
again calculated according to the model described in \cite{glaus2012}.
Due to the conjugate relationships between the variables in our model and those
of \cite{glaus2012}, the models are the same, subject to a slight
reformulation of the prior parameters.
\paragraph{Prior over $\btheta$\\}
The final part of our model is to specify some prior belief in the vector $\btheta$. To make our approximations tractable, it is necessary to use a conjugate prior, which in this case is a Dirichlet distribution
\begin{equation}
p(\btheta) = \frac{\Gamma(\hao)}{\prd m \Gamma(\amo)} \prd m \theta_m ^{\amo-1}
\end{equation}
where $\amo$ represents our prior belief in the values
of $\theta_m$ and $\hao = \sumover m \amo$. We use a weak but proper prior $\amo=1;\, m=0\ldots M$ which corresponds to a single ``pseudo-count" read (or read-pair) for each transcript.

\subsection{Approximate inference}
We are interested in computing the posterior distribution for the mixing proportions, $p(\btheta\given\bR,\bT) \propto {\sum_\bZ p(\bR\given \bT, \bZ)p(\bZ\given\btheta)p(\btheta)}$.
For very small datasets, it is possible to perform exact Bayesian inference in
this model, however for any realistically sized problem, exact inference is impossible due
to the combinatorial explosion of the number of possible solutions.
Our proposed solution is to use a collapsed version of Variational Bayes (VB).
VB involves approximating the posterior probability density of all the model parameters with another distribution $q$,
\begin{equation}
q(\btheta, \bZ) \approx p(\btheta, \bZ|\bR,\bT).
\end{equation}
The approximation is optimised by minimising the Kullback-Leibler (KL) divergence between $q(\btheta, \bZ)$ and $p(\btheta, \bZ|\bR,\bT)$~\citep{bishop2006pattern}. To make the VB approach tractable, some factorisations need to be assumed in the approximate posterior. In the case of the current model, we assume that the posterior probability of the transcript proportions factorises from the alignments:
\begin{equation}
q(\btheta,\bZ) = q(\btheta)q(\bZ).
\end{equation}
Further factorisations in $q(\bZ)$ occur due to the simplicity of the model, revealing $q(\bZ) = \prd n q(\bz_n)$.
We write the approximate distribution for $q(\bZ)$ using the parameters $\phi_{nm}$:
\begin{equation}
q(\bZ) = \prd n \prd m \phi_{nm}^{z_{nm}}.
\end{equation}
We need not introduce parameters for $q(\btheta)$ since it will arise implicitly in our
derivation in terms of $\phi$.

\paragraph{The objective function\\}
Approximate inference is performed by optimisation: the parameters of the
approximating distribution are changed so as to minimise the KL divergence.
Whilst the KL divergence is not computable, it is possible to derive a lower
bound on the marginal likelihood, maximisation of which minimises the KL
divergence~\citep[see e.g.][]{bishop2006pattern}.
Here we derive a lower bound which is dependent only on the parameters of
$q(\bZ)$, with the optimal distribution for $q(\btheta)$ arising implicitly
for any given $q(\bZ)$.
First we construct a lower bound on the conditional log probability of the reads $\bR$ given the transcript proportions $\btheta$ and the known transcriptome $\bT$:
\begin{equation}
\begin{split}
\ln p(\bR \given \bT, \btheta) &= \ln \int\! p(\bR\given \bZ, \bT) p(\bZ\given\btheta) \d \bZ\\
&\geq \E {q(\bZ)} {\ln p(\bR\given \bZ, \bT) + \ln p(\bZ\given\btheta) - \ln q(\bZ)}\\
&= \sumover n \sumover m \phi_{nm} \big(\ln p(\br_n\given T_m) + \ln \theta_m - \ln \phi_{nm}\big)\\
&= \mathcal L_1(\btheta) ,
\end{split}
\end{equation}
where the first line follows from Jensen's inequality in a similar fashion to standard VB methods.
We have denoted this conditional bound $\mathcal L_1(\btheta)$, which is still a function of $\btheta$.
In order to generate a bound on the marginal likelihood, $p(\bR\given \bT)$, we
need to remove this dependence on $\btheta$ which we do in a Bayesian fashion,
by substituting $\mathcal L_1(\btheta)$ into the following Bayesian marginalisation:
\begin{equation}
\begin{split}
p(\bR\given \bT) &= \int\! p(\bR\given \bT,\btheta) p(\btheta) \d \btheta\\
&\geq \int\! \exp\{\mathcal L_1(\btheta)\}p(\btheta) \d \btheta .
\end{split}
\end{equation}
Solving this integral and taking the logarithm gives us our final bound which equates to
\begin{equation}
\begin{split}
\ln p(\bR\given \bT) \geq \mathcal L = \sumover n \sumover m \phi_{nm} \big(\ln p(\br_n\given T_m) - \ln\phi_{nm} \big)\\
+ \ln \Gamma(\hao) -\ln \Gamma(\hao + N) -\sumover m \Big(\ln \Gamma(\amo)-\ln \Gamma(\amo + \hat\phi_m)\Big) ,
\end{split}
\label{eq:bound}
\end{equation}
where $\hat\phi_m = \sumover n \phi_{nm}$ and we also have that the approximate posterior distribution for $\btheta$ is a Dirichlet distribution with parameters $\amo + \hat\phi_m$. 

\subsection{Optimisation}
Having established the objective function as a lower bound on the marginal likelihood, all that remains is to optimise the variables of the approximating distribution $q(\bZ,\btheta)$. The dimensionality of this optimisation is rather high and potentially rather difficult. Optimisation in standard VB is usually performed by an EM like algorithm, which performs a series of convex optimisations in each of the factorised variables alternately.
In our formulation of the problem, we only need to optimise the parameters of the distribution $q(\bZ)$, which we do by a gradient-based method. Taking a derivative of \eqref{eq:bound} with respect to the parameters $\phi$ gives
\begin{equation}
\frac{\partial \mathcal L}{\partial \phi_{nm}} = \ln p(\br_n\given T_m) - \ln \phi_{nm} - 1 + \psi(\amo + \hat \phi_m) ,
\end{equation}
where $\psi$ is the digamma function. To avoid constrained optimisation we re-parameterise $\phi$ as $\gamma$:
\begin{equation}
\phi_{nm} = \frac{e^{\gamma_{nm}}}{\sum_{m'=1}^M e^{\gamma_{nm'}}}
\end{equation}
and it is then possible to optimise the variables $\gamma$ using a standard gradient-based optimiser.
\subsection{Geometry}
Information geometry concerns the interpretation of statistical objects in a
geometric fashion. Specifically, a class of probability distributions behaves as
a Riemannian manifold with curvature given by the Fisher information.
\cite{amari1998natural} showed that the direction of the steepest descent on
a such a manifold is given by the natural gradient:
\begin{equation}
\widetilde \nabla \mathcal L = G^{-1}\nabla \mathcal L \ ,
\end{equation}
where $G$ is the Fisher information matrix. Since we are performing optimisation of the distribution $q(\bZ)$, we can make use of the natural gradient in computing a search direction \citep{honkela2010approximate}. For our problem, we assume that the $N \times M$ matrix $\bZ$ has been transformed into a $NM$ vector, and the Fisher information corresponding to $\gamma_{nm}$, $\gamma_{n'm'}$ is given by
\begin{equation}
G[m,n,m',n'] = \left\{ \begin{array}{rl}
\phi_{nm} - \phi_{nm}^2,&\mbox{ if $n=n'$ and $m=m'$ } \\
-\phi_{nm}\phi_{nm'},&\mbox{ if $n=n'$ but $m\neq m'$ } \\
0,&\mbox{ otherwise.}
\end{array} \right.
\end{equation}
We note that this structure is block-diagonal, and that each block can be easily inverted using the Sherman-Morrison identity, giving an analytical expression for $G^{-1}\nabla \mathcal L$, and thus making the natural gradient very fast to compute (see \citet{hensman2015tpami} for more details). One can draw comparisons with a Newton method, where $G$ would be replaced with a Hessian, though in the proposed case the system is much cheaper to compute. 

The optimisation of the variational parameters then proceeds as follows.
Following random initialisation, a unit step is taken in the natural gradient
direction.
Subsequent steps are subject to conjugate gradients \citep{honkela2010approximate}. If the conjugate gradient step should fail to improve the objective we revert to a VBEM update, which is guaranteed to improve the bound. For more details, see \cite{hensman2012fast}.
\subsection{Truncation}
The optimisation described above has $N \times M$ free parameters for optimisation, one to align each read to each transcript. However, for most read-transcript pairs, $p(\br_n\given T_m)$ will be negligibly small. We follow \cite{glaus2012} in truncating the values of $p(\br_n\given T_m)$ to zero for reads which do not suitably align. Examining the objective function \eqref{eq:bound} we see that we can also set $\phi_{nm}$ to zero for these truncated alignments (using the convention that $0\ln(0) = 0$) and thus also $\gamma_{nm} = -\infty$ for the same. This truncation dramatically reduces the computational load of our algorithm, reducing the dimensionality of the optimisation space as well as reducing the number of operations needed to compute the objective.

\subsection{The approximate posterior}
Having fitted our model, we may wish to propagate the posterior
distribution through a second set of processing, for example to
identify differentially expressed transcripts as in BitSeq stage 2
\citep{glaus2012}. Whilst it may be desirable to solve both stages
together in a Bayesian framework, the size of the problem generally
forbids this, therefore we propose the use of either a moment-matching
or sampling procedure to propagate $q(\btheta)$ through further
analysis. The approximate posterior $q(\btheta)$ is a Dirichlet
distribution, whose marginals have the following useful properties:
\begin{align}
\E {} {\theta_m} &= \frac{\amo + \hat\phi_m}{\hao + N} ,\\
\text{var} [\theta_m] &= (\amo + \hat\phi_m)(\hao + N - \amo - \hat\phi_m) C , \\
\text{cov} [\theta_m, \theta_m'] &= -(\amo + \hat\phi_m)(\alpha_{m'}^o + \hat\phi_{m'}) C ,
\end{align}
with $C = (\hao+ N)^{-2}(\hao + N + 1)^{-1}$.
This approximate posterior is somewhat inflexible, in that it cannot express
arbitrary covariances between the transcripts. This arises from the factorising
assumption amongst the assignment of reads to transcripts: reads are assigned
independently in the variational method and their dependence cannot be modelled.
This is reflected in the results section where we show empirically that the VB
approximation leads to an underestimation of the variance. Nonetheless, this
simplifying assumption leads to very accurate expression estimates much faster
than MCMC.

\begin{figure*}[ht]
\centering
\begin{tabular}{c}
\includegraphics[width=0.95\textwidth]{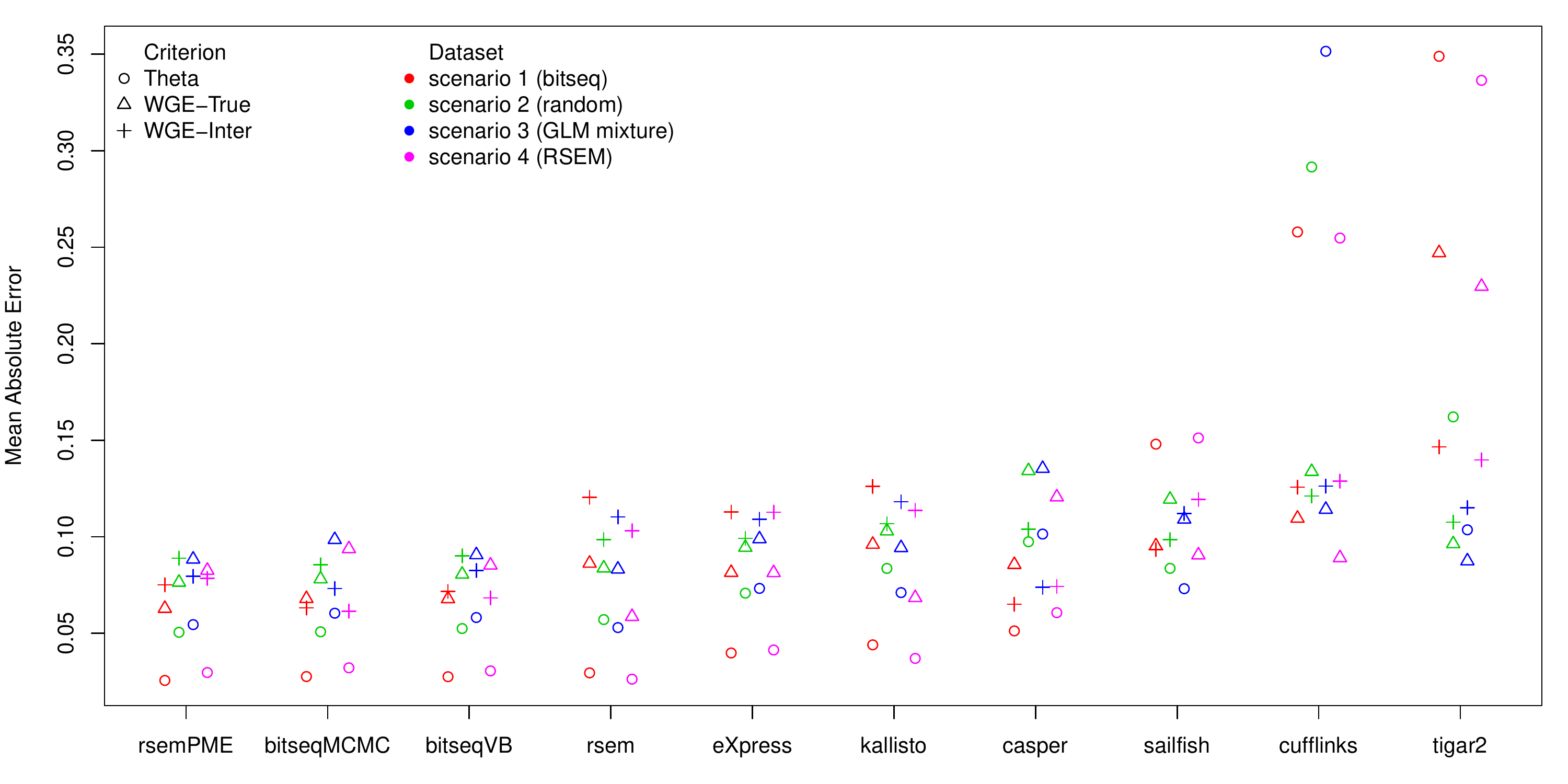}\\
\end{tabular}
\vspace{0.0em}
\caption{Ranking of methods for five replicates of simulated RNA-seq reads.
WGE-Inter: inter-replicate consistency of within gene estimates, WGE-True:
within gene estimates compared to the true values and Theta: estimated relative
transcript expression compared to the true values. Scores have been nomrmaised to unity per dataset. Alternatives normalizations do not change the ranking (see supplementary material).}

\label{fig:sim1}
\end{figure*}

\section{Results and Discussion}
\label{sec:results}

The proposed BitSeqVB algorithm was compared to Cufflinks \citep{trapnell2010}, RSEM as well as the corresponding MCMC sampler RSEM-PME \citep{Li2011}, BitSeqMCMC \citep{glaus2012},  eXpress~\citep{roberts2013streaming}, Casper~\citep{rossell2014},  Sailfish~\citep{patro2014sailfish}, Tigar2~\citep{tigar2} and Kallisto \citep{kallisto}. We note that both MCMC samplers (RSEM-PME and BitSeqMCMC) use similar collapsed Gibbs algorithms but are initialised differently: RSEM-PME starts from the ML solution found by RSEM while BitSeqMCMC starts from a random initialisation and therefore requires more interations to find a good solution.

We used two
main ways for benchmarking: analysis on synthetic data allowed comparison with a known ground
truth under a variety of generative scenarios; analysis on high-quality replicated human data focussed on inter-replicate consistency following the evaluation of \cite{rossell2014}. We find BitSeqVB to have excellent inter-replicate 
consistency and accuracy, closely approximating the original MCMC 
algorithm, while also being competitive with other methods in terms of run-time. We subsequently analyse in more detail the approximation to the posterior used in the BitSeqVB method. For comparison with other methods, we used default settings where appropriate: both MCMC sampling methods use 1000 posterior samples as default. However, this number refers to effective samples \citep{gelman2003bayesian} in BitSeqMCMC and not to single iterations as in RSEM-PME.  We turned off creating of unecessary output files in RSEM. The experiments were conducted on a 4 core workstation. All the details of the experiments can be found at the aforementioned url.

\subsection{Inference accuracy on synthetic data}\label{sec:sim}

RNA-seq reads from $M = 48009$ transcripts of the UCSC/hg19 transcriptome annotation \citep{ucsc} were simulated using the Spanki software \citep{spanki}.
The expression is evaluated in three different measures: transcript expression accuracy (Theta), transcript within-gene relative
proportion accuracy (WGE-True) and inter-replicate consistency (WGE-Inter).

A ground truth was generated using four different models of transcript expression, according to the following scenarios: 
\begin{enumerate}
\item estimated expression levels from real data using BitSeqMCMC ($\approx 56$ million reads per replicate)
\item randomly selected expression levels according to a uniform distribution defined on the set $(10,200)$ ($\approx 7.8$ million reads per replicate)
\item a high-dimensional mixture of Poisson Generalized Linear models, which was recently used to model the heterogeneity in RNA-seq datasets \citep{poisGLM} ($\approx 5.5$ million reads per replicate)
\item estimated expression levels from real data using RSEM ($\approx 18$ million reads per replicate)
\end{enumerate}
For each scenario five replicates are generated according to a Negative Binomial model. Full details of the four scenarios are described in the Supplementary Material. Finally, the resulting reads-per-kilobase (RPK) values were fed into Spanki. Next, the simulated reads were aligned to the reference annotation using
Bowtie2 and/or Tophat2. In particular, BitSeq, RSEM, eXpress and Tigar require
transcriptomic alignments so Bowtie2 (version 2.0.6) \citep{Langmead2012} was used, while Cufflinks and Casper work
with genomic alignments using Tophat2 and Bowtie2. On the other hand, Sailfish and Kallisto
produce their own alignments using $k$-mers mapping and pseudo-alignments, respectively. The corresponding mapping
rate for genomic or transcriptomic alignments was $96\%$. The same amount of reads pseudo-aligned when using Kallisto, whilst Sailfish
mapped a smaller portion of $k$-mers ($\approx 63\%$).

\begin{figure}[t]
\centering
\begin{tabular}{c}
\includegraphics[width=0.75\textwidth]{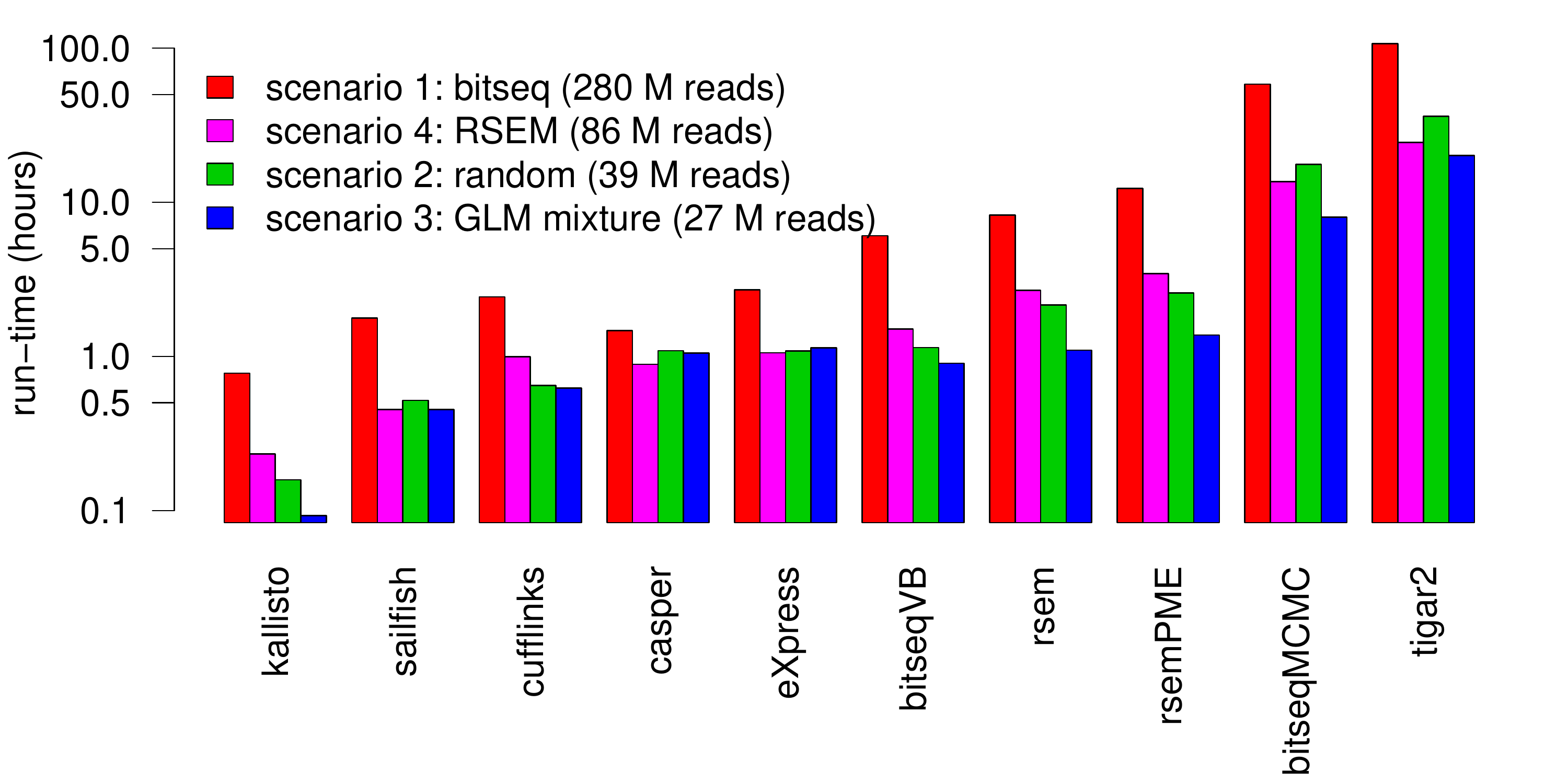}\\
\end{tabular}
\vspace{0.0em}
\caption{Run-time in hours (log-scale) for 4 synthetic data samples with 5 replicates per sample. The total number of simulated reads is shown in parenthesis.}
\label{fig:spankiTime}
\end{figure}

Figure \ref{fig:sim1} displays the mean absolute error (MAE) according to the three
criteria, after performing the following normalization: $$\sum_{m \in \mbox{methods}}\mbox{MAE}^{(c)}_m = 1, $$ $\forall c \in\{\mbox{Theta},\mbox{WGE-Inter},\mbox{WGE-True}\}$, in order to make all criteria equally weighted for each scenario. Moreover, 
the ``Theta'' and ``WGE-True'' metrics were averaged across the five replicates, while
``WGE-Inter'' was averaged across all ten combinations of pairs of replicates.
The methods were ranked with respect to their average across the three criteria.
RSEM-PME, BitSeqMCMC and BitSeqVB are ranked as best when considering all three criteria. RSEM has
similar accuracy in terms of the ground truth expression (Theta and WGE-True) but has lower inter-replicate consistency (WGE-Inter). Conversely, Casper achieves good performance with respect to inter-replicate consistency (WGE-Inter) but is less accurate in comparison to the ground truth values (WGE-True and Theta). The ranking of
methods with respect to run-time is shown in Figure \ref{fig:spankiTime}. Note that the run-time calculation excludes the
alignment procedure, but includes all other computations (including computing alignment probabilities in BitSeq's case). An exception is made for Sailfish and Kallisto, where alignment is not required, making these by far the fastest methods. Timings which include the time required for alignment are provided in Supplementary Figure \ref{fig:sim1-mapTimes}.

The plots of inter-replicate consistency between pairs of replicates are shown in the supplementary material (Figures 2, 4, 6 and 8). As seen there, Kallisto, RSEM, Sailfish, Tigar2, Cufflinks and eXpress, produce estimates close to the boundary of the parameter space. This is also obtained for RSEM-PME except for scenario 2. This behaviour is avoided when using BitSeqMCMC, BitSeqVB and Casper. 


The accuracy of BitSeqVB is very close to the two sampling methods
BitSeqMCMC and RSEM-PME, but it is consistently faster that these
approaches, being about 10 times faster than BitSeqMCMC and 2 times faster than RSEM-PME on average (RSEM-PME is significantly faster than BitSeqMCMC because is uses many fewer iterations of MCMC). BitSeqVB has similar speed to the eXpress method in most cases whilst exhibiting much better accuracy.

We conclude that the proposed VB algorithm is competitive in speed while 
exhibiting both high accuracy and good inter-replicate consistency.

\subsection{Replicate consistency in human data}\label{sec:humanData}

A recent study \citep{rossell2014} used the mean absolute error between
pairs of replicates of the same ENCODE experiment in order to assess
the accuracy of transcript expression estimation methods. For this
purpose, the relative within gene expression estimates are used (WGE-Inter). Here, we
provide an extended version of this analysis in order to benchmark
against BitSeqMCMC and seven other methods.

\begin{figure}[ht]
\centering
\begin{tabular}{c}
\includegraphics[width=0.95\textwidth]{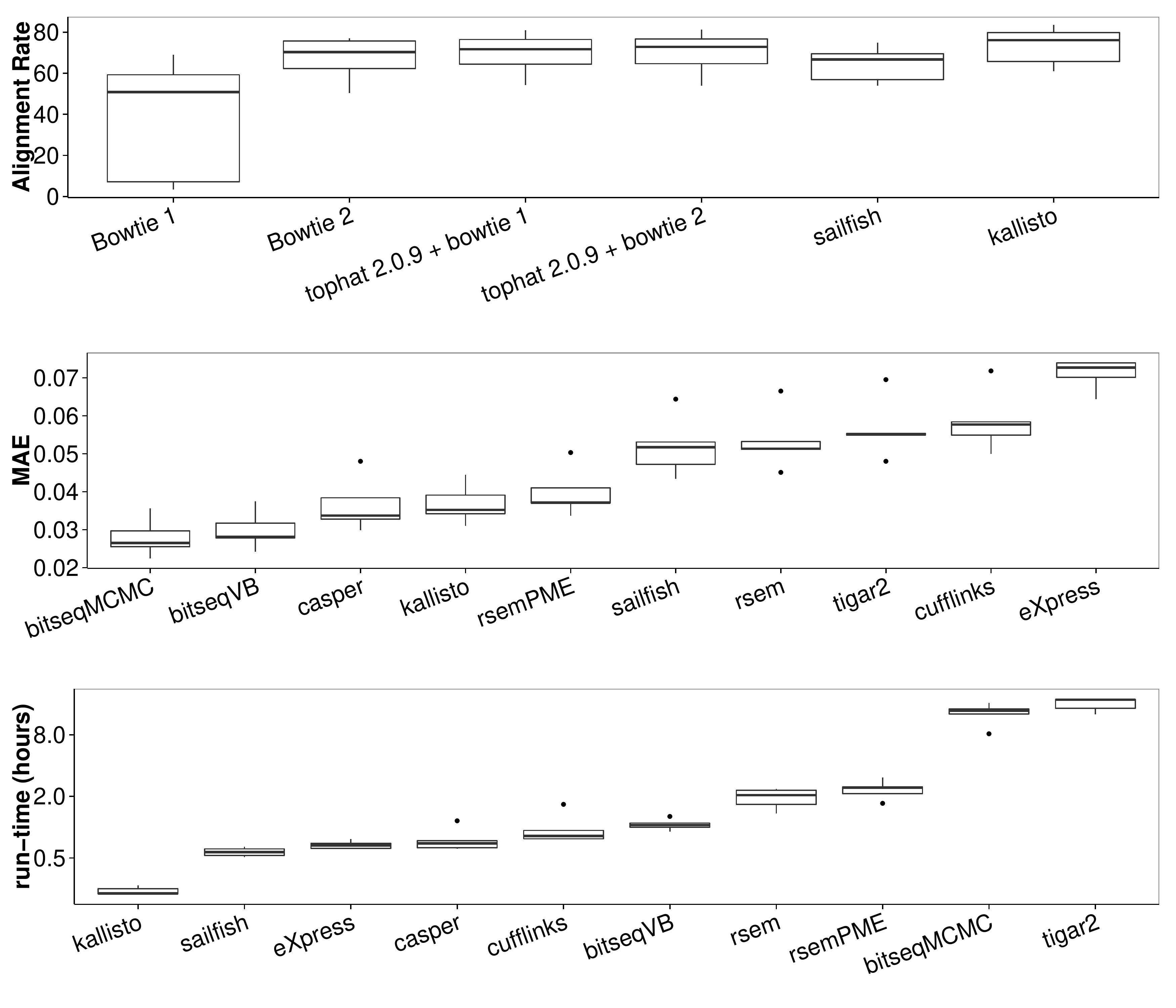}
\end{tabular}
\vspace{0.0em}
\caption{Five ENCODE pairs of replicates. (a): Alignment rates for transcriptome
mapping (Bowtie1 and Bowtie2), genome mapping (Tophat 2.0.9 with Bowtie1 and
Bowtie2), k-mers mapping (Sailfish) and pseudo-alignments (Kallisto). (b): Ranking of methods in terms of the
Mean Absolute Error. (c): Run-time in hours (log-scale) with 24.6M (mapped) reads per sample.}
\label{fig:methodsRanking}
\end{figure}

In total, five ENCODE datasets~\citep{Tilgner2012} consisting of $2\times 76$ bp reads were selected,
corresponding to the following pairs of replicates: (SRR307897, SRR307898),
(SRR307901, SRR307902), (SRR307907, SRR307908), (SRR307911, SRR307912), (SRR307915, SRR307916). 
All methods were applied assuming the same UCSC/hg19 transcriptome annotation as in the previous section. According to the alignment rates
shown in Figure \ref{fig:methodsRanking}(a), all methods work with almost the
same number of mapped reads when Bowtie2 is used. This is not the case for
Bowtie1 which for some reason fails on this data set.

\begin{figure*}[t!]
\centering
\includegraphics[width=0.95\textwidth]{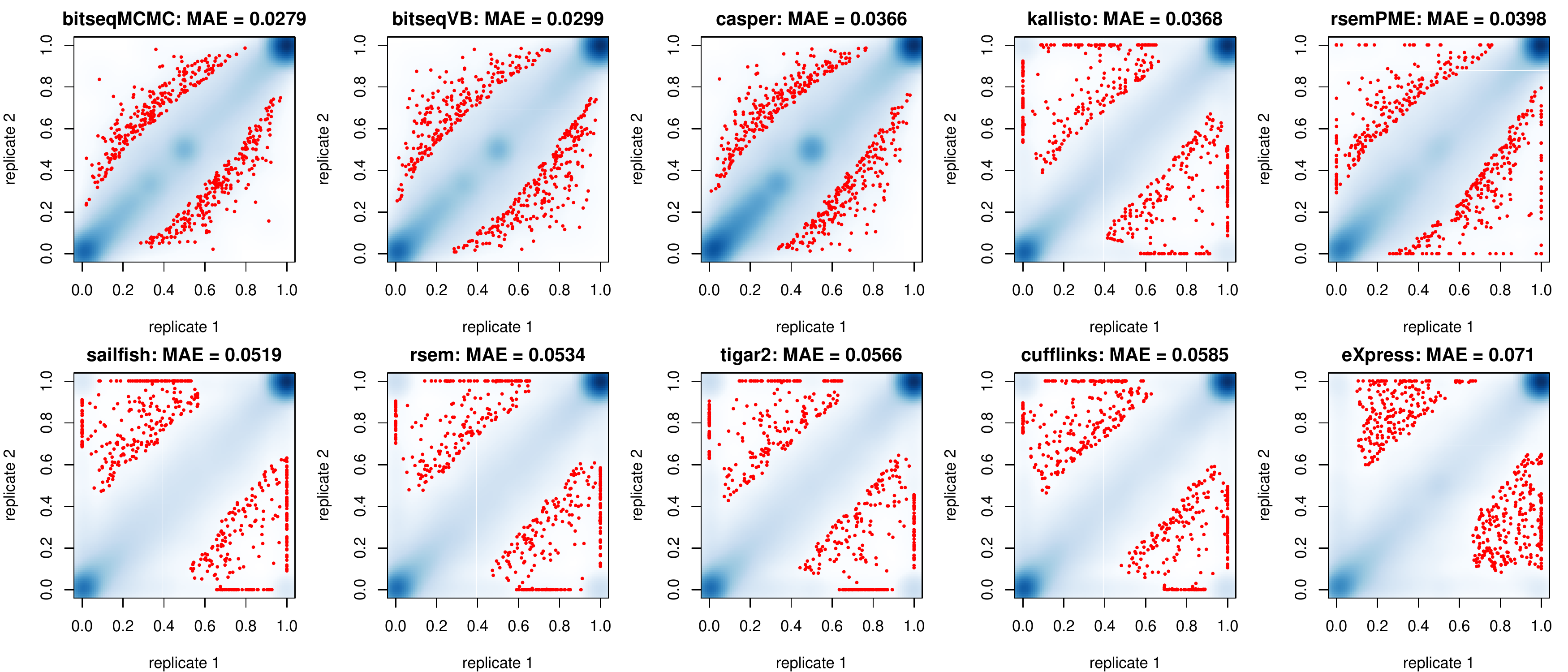}\\
\caption{Scatterplots of within gene estimates for one pair of replicates
(SRR307907 and SRR307908) from the ENCODE data. The blue color corresponds to a
smoothed color density representation of the scatterplot.}
\label{fig:wge}
\end{figure*}

Figure \ref{fig:methodsRanking}(b) illustrates the ranking of methods in terms
of the MAE criterion, averaged across the five datasets. We conclude that
BitSeqMCMC has best inter-replicate consistency, closely followed by BitSeqVB, while
Casper comes next. Sailfish, RSEM, Tigar2 and Cufflinks exhibit almost two
times larger MAE, while eXpress is almost 2.5 times worse according to this 
measure. Based on these five
samples there is a partial order: $\mbox{BitSeqMCMC} \succ \mbox{BitSeqVB} \succ
\{\mbox{Casper, Kallisto}\}\succ \mbox{RSEM-PME} \succ \{\mbox{RSEM, Sailfish}\} \succ  \{\mbox{Cufflinks, Tigar2}\} \succ
\mbox{eXpress}$, where $\succ$ denotes ``is better in every experiment''. Excluding
BitSeq (MCMC and VB) and Casper, we see that many methods produced
estimates close to the boundary of the parameter space, as seen in Figure
\ref{fig:wge}. This means that many transcripts are estimated as weakly or
non-expressed in one replicate while being highly expressed in the other. This problem 
appears to affect methods using Maximum Likelihood estimation (RSEM, Sailfish, Cufflinks, eXpress) or Bayesian methods using a very weak prior (Tigar2). Casper ensures consistency with a strong prior, but this may degrade the accuracy of absolute estimates relative to BitSeq because of stronger regularisation. We note that Casper uses MAP parameter estimation, finding the mode of the posterior distribution, while the BitSeq methods estimate the mean of the posterior distribution. Using the posterior mean may avoid spurious values where the mode is a long way from the mass of the posterior without the need for an overly strong prior. Finally, note that the coherency of inter-replicate consistency estimates in our simulation study (supplementary Figures 2,4,6 and 8) with the one reported here.

The run-time for each method is displayed in Figure
\ref{fig:methodsRanking}(c). BitSeqVB is comparable to the fastest methods (except for Kallisto which is by far the fastest method) while being
ranked as second in terms of the MAE criterion. We conclude that BitSeqVB 
offers perhaps the best trade-off in accuracy and runtime
on these datasets. 

Finally, we mention that the BitSeqMCMC performance here is in stark contrast with
the performance reported in \cite{rossell2014}. The reason for this is that in
\cite{rossell2014} reads were aligned using Bowtie1 whereas we are using Bowtie2. As seen in Figure
\ref{fig:methodsRanking}(a), Bowtie1 can exhibit very low alignment rates
for these samples. Interestingly, this behaviour is not
present when Bowtie1 is combined with Tophat for genome mapping. The
low alignment rates of Bowtie1 means that methods have available only a tiny fraction
of the useful data, leading to less accurate results. This explains the weak
agreement of same transcript estimates between pairs of replicates reported
for BitSeq in \cite{rossell2014} and is a reminder that it is very
important to check the alignment rates.

\subsection{Analysis of the Variational Bayes approximation}

To examine the properties of the variational approximation, we focussed on
ENCODE dataset SRR307907~\citep{Tilgner2012}. This contained 30.8 million reads, each 76bp. The reads were again mapped to the same UCSC/hg19 reference transcriptome resulting in
23.7 million mapped reads. 

\begin{figure}[!tpb]
\centering
{\includegraphics[width=0.95\textwidth]{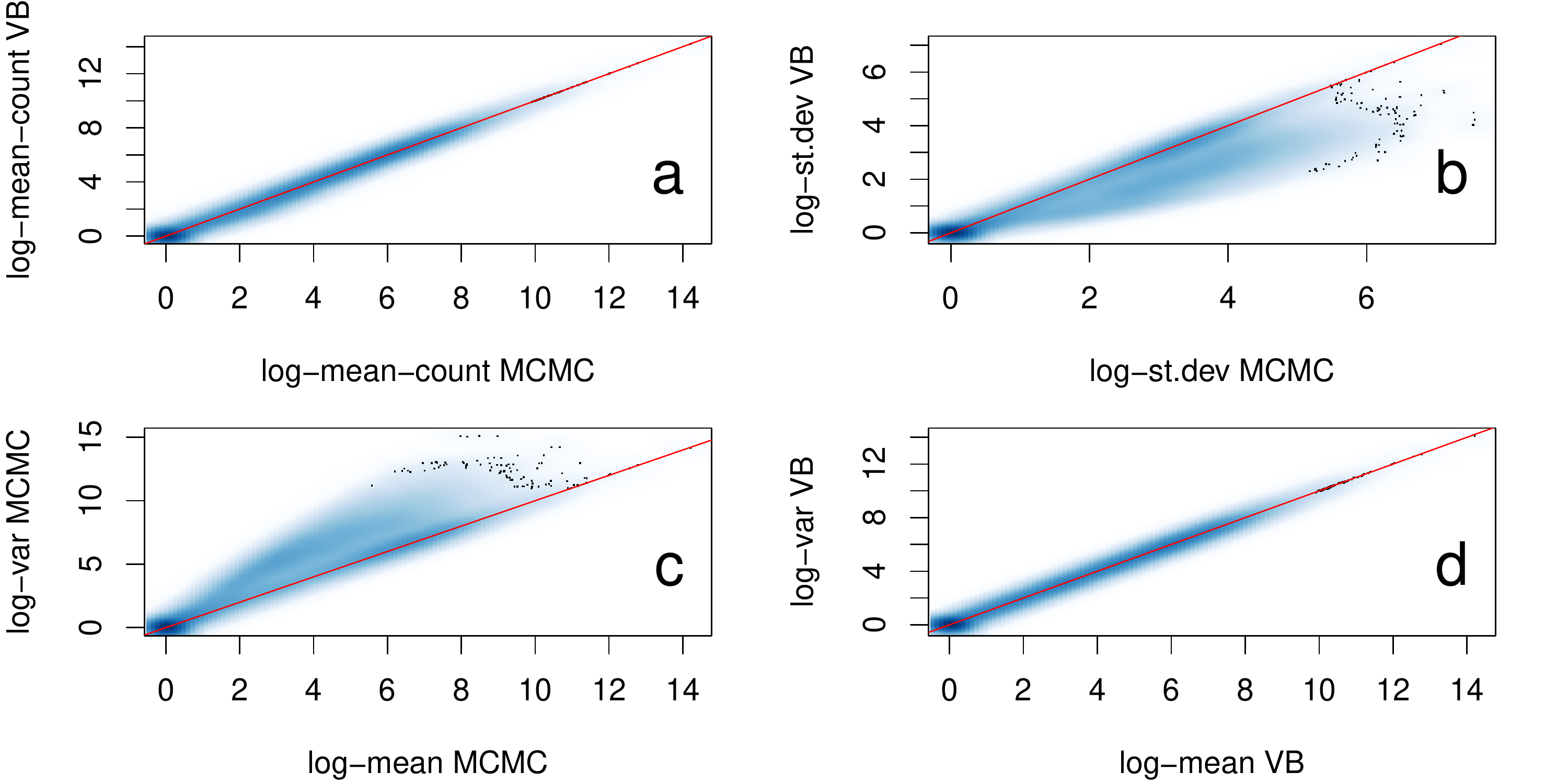}}
\vspace{0.0em}
\caption{A comparison of the first two moments of the approximate posterior
expression in counts per transcript:
(a) posterior mean ($R^2$ correlation is $0.999$)
(b) posterior standard deviation: the VB method significantly under-estimates the
posterior variance ($\sigma^2$).
(c), (d) posterior mean-variance relationship in MCMC and VB respectively.
Shading represents the number of transcripts in each region.
}\label{fig:mcmcCorrelation}
\end{figure}

Our main potential concern in using the VB method is the
quality of approximation to the posterior. Figure \ref{fig:mcmcCorrelation}(a)
shows a comparison of the variational posterior with a ground truth computed by
MCMC with a very large sampling time.  We conclude that the VB method consistently provides very accurate estimates of the posterior mean across the whole range of expression levels.
The estimates of posterior variance are less consistent and for a fraction of
transcripts the variances are underestimated (Figure \ref{fig:mcmcCorrelation}(b)).  It appears that VB only estimates the Poisson variance associated with random sampling of reads
(Figure \ref{fig:mcmcCorrelation}(d)), whereas the true posterior variance is larger for some
transcripts due to the uncertainty in assigning multi-mapping reads (Figure \ref{fig:mcmcCorrelation}(c)). If estimation of the expression level is all that is required,
then it would seem that the VB method suffices.  However, downstream methods
which make use of uncertainty in the transcript quantification (such as the
differential expression analysis proposed in BitSeq stage 2~\citep{glaus2012}) may suffer from the poor
approximation in terms of posterior variance. This can potentially be addressed by augmenting 
the VB method with a
more accurate approximation as done in a recent study that proposed a new VB algorithm with
improved variance estimates and a tighter lower bound on the
log-marginal likelihood \citep{papastamoulis2014improved}.

\subsection{Convergence comparison}

We further investigate convergence properties of MCMC and VB in terms of mean
expression.  
RNA-seq data was obtained from ENCODE experiment SRX110318, run SRR387661, generating 124.8 million 76 bp read-pairs. We mapped the reads using Bowtie 2 to a reference transcriptome using 8713 transcripts of chromosome 19 from Ensembl human cDNA, release 70 \citep{ensembl}.

%

As the true expression levels are unknown, we used a
long run of MCMC as the ground truth for mean expression estimates.  Running the
inference methods for a certain number of iterations, we record the run time and
calculate Root Mean Square Error (RMSE) of estimated expression.
\begin{figure}[!tpb]
\centering
\includegraphics[width=0.75\textwidth]{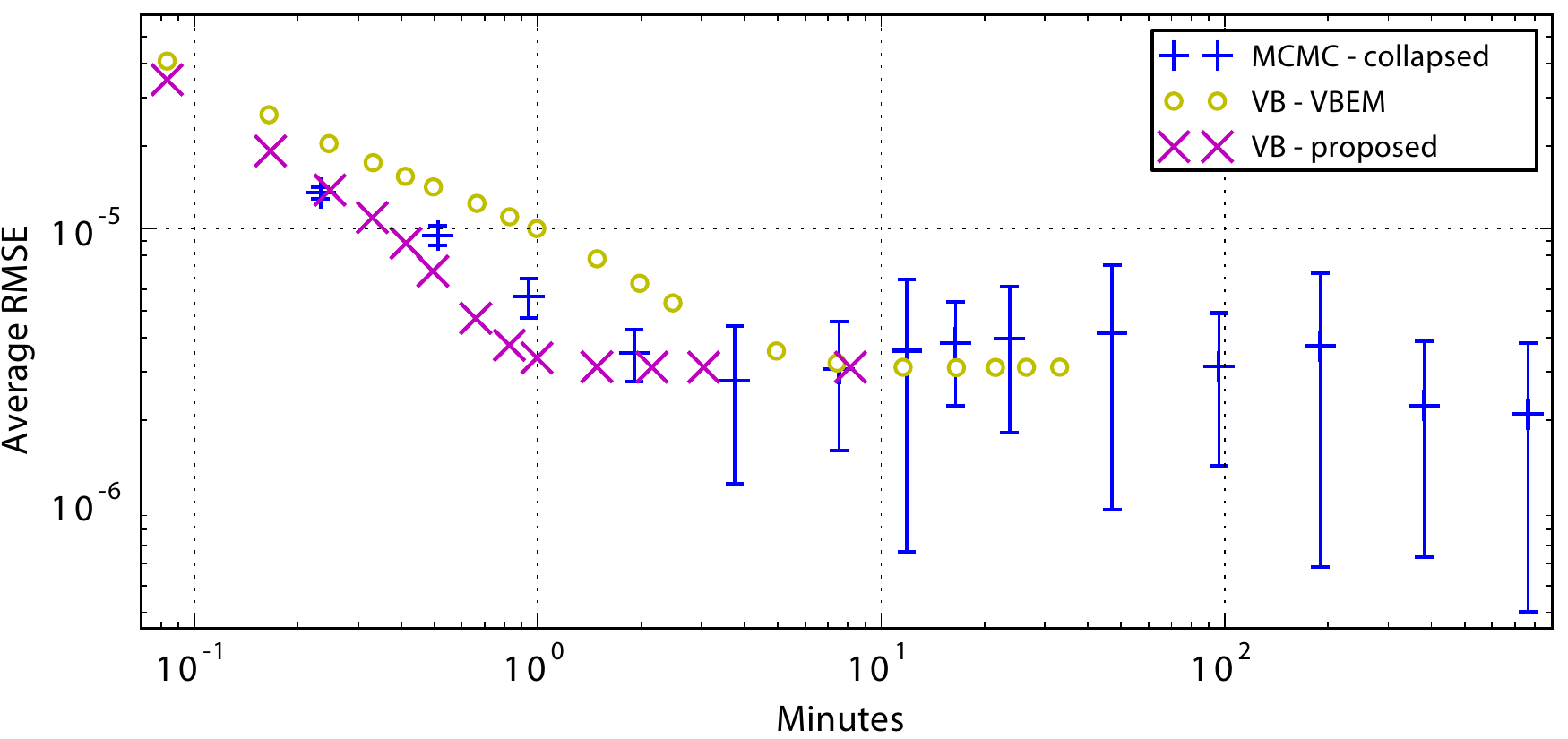}
\vspace{1.5em}
\caption{Convergence comparison of Collapsed MCMC with standard VB algorithm
and VB with Fletcher-Reeves conjugate gradient optimisation.  Expression
estimates obtained by very long run of MCMC are used as a ground truth and
average root mean square error over 10 runs was calculated, two standard
deviations are used as error bars. The VB methods with several randomised
initial conditions showed negligible differences in convergence.
}\label{fig:simConvergence}
\end{figure}
The convergence of our variational method (BitSeqVB) and the original Gibbs sampling procedure (BitSeqMCMC) is shown in Figure \ref{fig:simConvergence}.  We also include a standard implementation of VB (similar to \citet{nariai2013tigar}) but using
the BitSeq model (denoted VBEM).  It is straightforward to derive this
algorithm from our VB algorithm derivation since standard VBEM is obtained as a special
case of steepest descent VB learning~\citep{hensman2012fast}.
Our implementation of VB converges first in about 2 minutes.  Surprisingly,
some runs of collapsed MCMC converge to better estimates even faster than
standard VB, which takes around 10 minutes.  However, as MCMC is a stochastic
method, an estimate that is consistently better than the results obtain by VB
is only obtained after 900 minutes.

\section{Conclusion}

We have presented a new Variational Bayes method for inference of transcript
expression from RNA-seq data. Building on previous work in BitSeq, we have
presented a fast approximate inference method. The mean of the posterior
distribution of expression levels was very well estimated in substantially less
time than the original MCMC algorithm. The method is therefore suitable when
point estimates of expression are sufficient, especially if time and
computational resources are limited. We have compared both the original BitSeq
algorithm and our new method with the majority of available methods for
transcript expression estimation and conclude that BitSeqVB is highly
competitive both in terms of expression estimation and run-time. We also note
that an existing VBEM algorithm implementation, TIGAR, does not provide a
significant improvement over Gibbs sampling in terms of computational time in
our examples, as well as having a very high memory requirement.

The newest method considered here, Kallisto, is found to be extremely
fast and perform with very good accuracy compared to other maximum
likelihood approaches. This speed-up is achieved through avoiding full
alignment and simplifying the likelihood computation through using a
pseudo-alignment approach. However, the method still produces estimates
at the boundary in our between-replicate comparisons similar to all
maximum likelihood methods. It would therefore be very interesting to
apply a Bayesian algorithm, such as the fast VB method proposed here,
using the same likelihood model as Kallisto.

Finally, we suggest some areas for future development. The fast and consistent
convergence of the VB method makes it useful for quick examination of the data
before the Gibbs sampler is run. Further, since it provides an excellent
approximation to the mean of the posterior, it could be used to e.g. reduce the
burn-in time for the Gibbs sampler, or as the initial stage of a more
sophisticated approximating technique, as in \cite{papastamoulis2014improved}.

\section{Supplementary Data}
We provide details of the simulation study and additional Figures as Supplementary Material which is available online.

\paragraph{Funding} JH was supported by an MRC fellowship and BBSRC award
BB/H018123/2, PP and MR by BBSRC award BB/J009415/1 and MR by EU FP7 award ``RADIANT" (grant no. 305626). PG was supported by the Engineering and Physical Sciences Research Council [EP/P505208/1].
AH was supported by the Academy of Finland [259440 to AH].

\paragraph{Acknowledgements} We thank three anonymous reviewers for useful comments which have greatly improved the paper and Lior Pachter for his blog comments on a preliminary version of this paper which led us to include more realistic data simulation scenarios.

\appendix

\section{Simulation study details}

We describe the generative process of the RPK values used in the simulation section of the manuscript. 
Let
$\mathcal{NB(\mu,\phi)}$ denote the Negative Binomial distribution, with mean
equal to $\mu$ and variance equal to $\mu + \mu^2/\phi$, $\mu\geqslant 0$,
$\phi > 0$. Denote by $\mbox{RPK}_{jm}$ the RPK value for transcript $m =
1,\ldots,M$ at replicate $j=1,\ldots,J$, where $J$ and $M$ denote the number of
replicates and the total number of transcripts, respectively. 

\subsection{Scenario 1: BitSeq estimates from real data}

\begin{figure*}[t]
\centering
\includegraphics[width=0.9\textwidth]{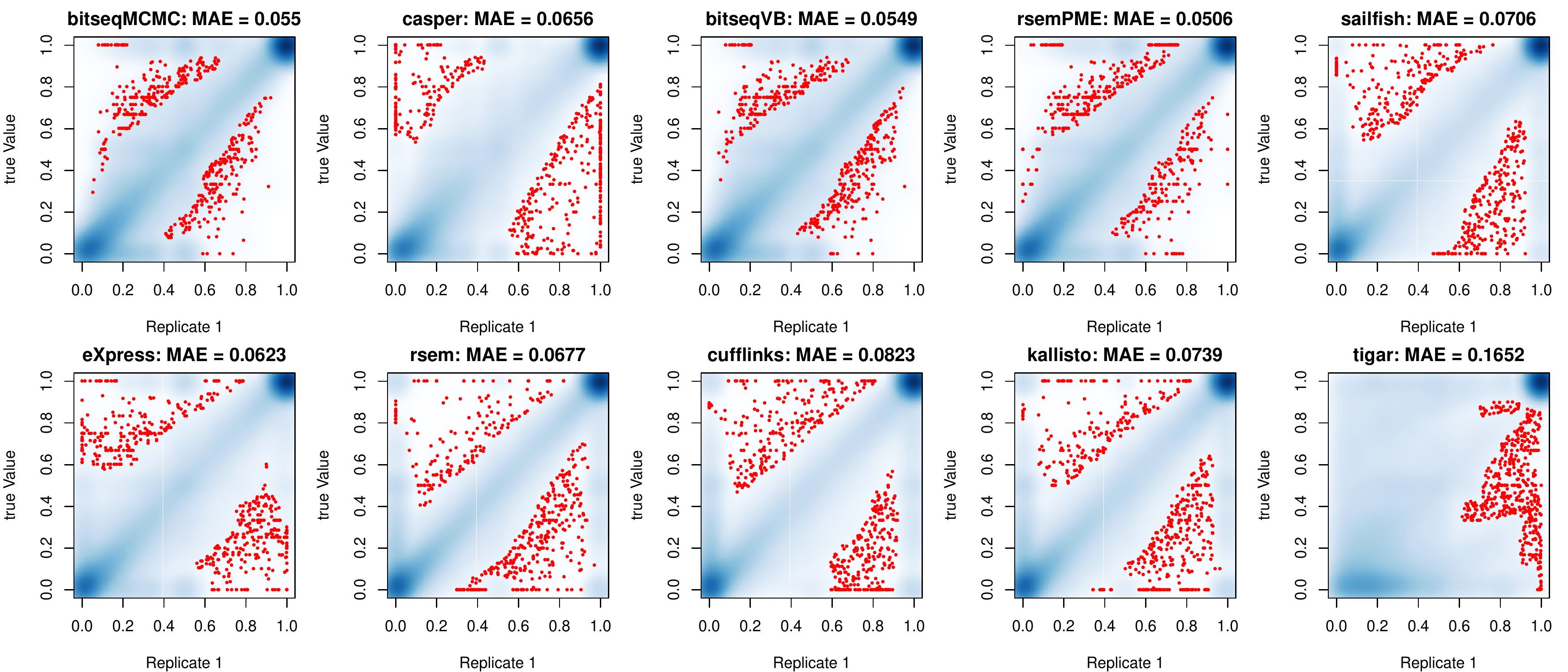}\\
\caption{Scenario 1: Scatterplots of within gene estimates for one replicate of the
simulated data versus the true values. The blue color corresponds to a smoothed
color density representation of the scatterplot and the red color emphasizes
points from the lowest regional densities.}
\label{fig:bitseq-spanki1}
\end{figure*}
\begin{figure*}[ht]
\centering
\includegraphics[width=0.9\textwidth]{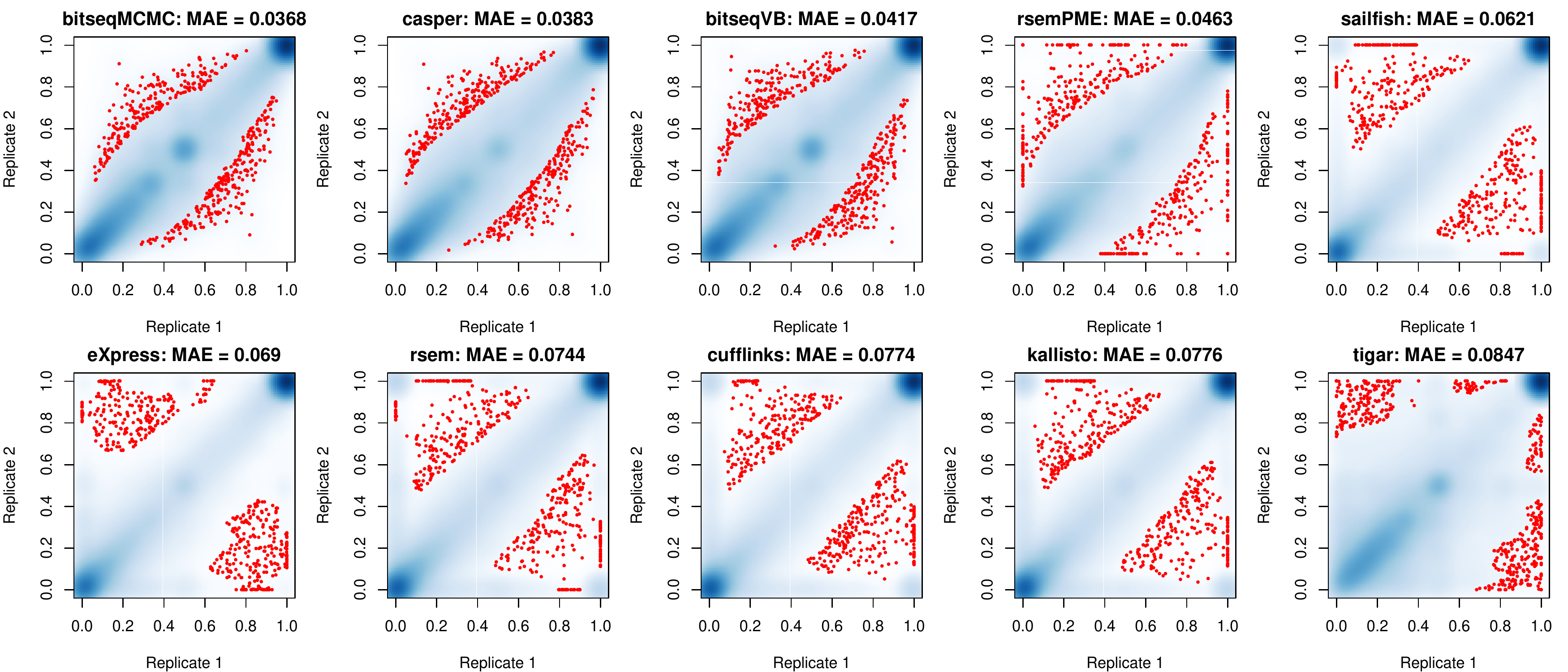}\\
\caption{Scenario 1: Scatterplots of within gene estimates for one pair of replicates from
the simulated RNA-seq reads. The blue color corresponds to a smoothed color
density representation of the scatterplot and the red color emphasizes points
from the lowest regional densities.}
\label{fig:bitseq-spanki2}
\end{figure*}

We used human data (SRR307907) from the ENCODE project in order to capture the dynamics of realistic RNA-seq datasets. BitSeq (MCMC) was used to estimate the relative expression levels of $M=48009$ transcripts. Next, the resulting estimates was used as input to generate the baseline mean. Reads were simulated according to the following generative process: 
\begin{eqnarray*}
\mbox{RPK}_{jm} &\sim& \mathcal{NB}(\widehat\mu_{m},50),\quad m=1,\ldots,M, j = 1,\ldots,5.
\end{eqnarray*}
with $\widehat\mu_m$, $m=1,\ldots,M$ denoting the corresponding estimates of RPK values according to BitSeq MCMC. This resulted in $56$ million paired-end reads  of 76 base-pairs per
replicate, almost $280$ million reads in total.  

Figures \ref{fig:bitseq-spanki1}
and \ref{fig:bitseq-spanki2} illustrate the scatterplots of within gene estimates for
the first replicate versus the true values and the second replicate estimates,
respectively. The plots are ordered according to the inter-replicate consistency Mean Absolute Error, as shown in Figure \ref{fig:bitseq-spanki2}. Note that BitSeqMCMC, BitSeqVB and Casper are the only methods which
avoid extreme outliers on the boundary of the inter-replicate consistency graphs.

\subsection{Scenario 2: random selection}

A ground truth was generated using randomly selected levels of transcript expression, and
the resulting reads-per-kilobase (RPK) values were fed into Spanki. Let $\mathcal U(\alpha,\beta)$ denotes the uniform distribution defined on the set $(\alpha,\beta)$, $\alpha<\beta$. Reads were simulated according to the following generative process:
\begin{eqnarray*}
\mu_m &\sim& \mathcal U(10,200),\quad m=1,\ldots,M\\
\mbox{RPK}_{jm} &\sim& \mathcal{NB}(\mu_{m},20),\quad m=1,\ldots,M, j = 1,\ldots,5.
\end{eqnarray*}
This resulted in $15.8$ million paired-end reads  of 76 base-pairs per
replicate, almost $80$ million reads in total.  

\begin{figure*}[t]
\centering
\includegraphics[width=0.9\textwidth]{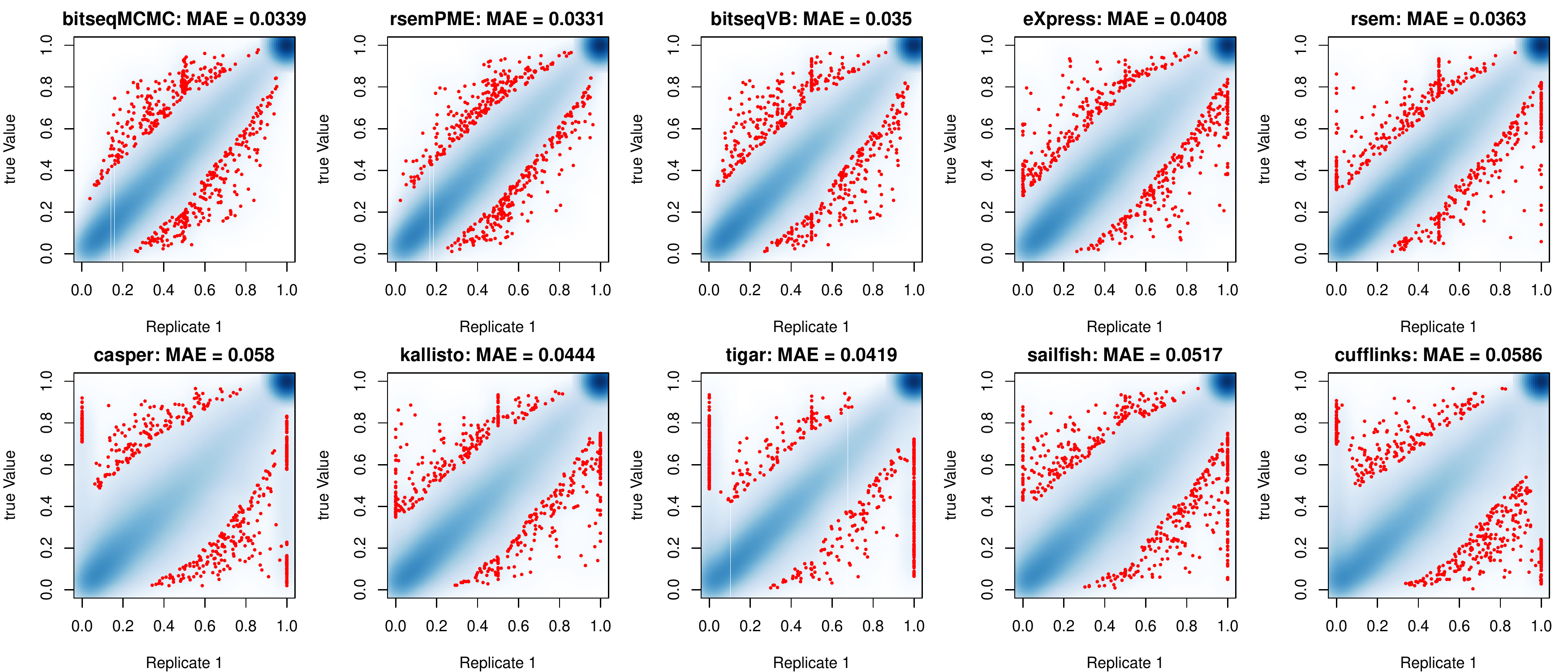}\\
\caption{Scenario 2: Scatterplots of within gene estimates for one replicate of the
simulated data versus the true values. The blue color corresponds to a smoothed
color density representation of the scatterplot and the red color emphasizes
points from the lowest regional densities.}
\label{fig:spanki1}
\end{figure*}

\begin{figure*}[t]
\centering
\includegraphics[width=0.9\textwidth]{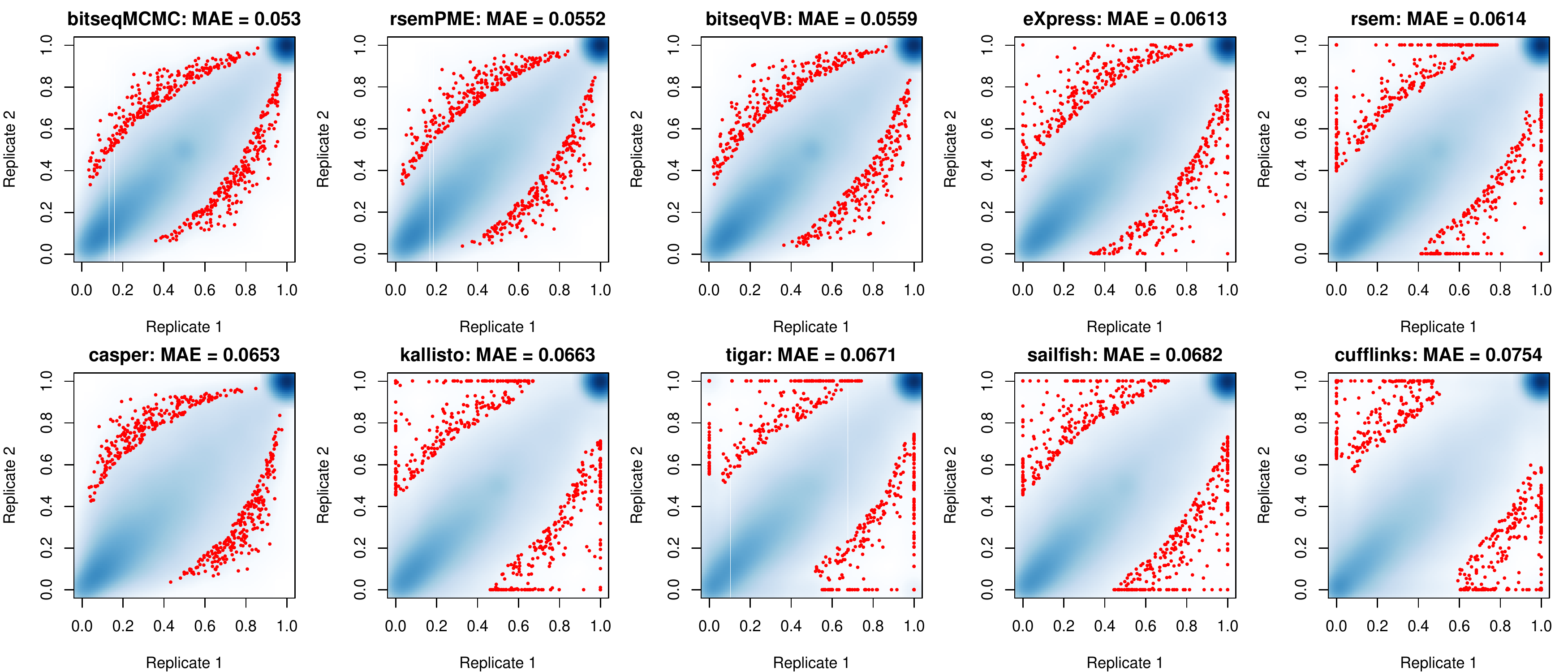}\\
\caption{Scenario 2: Scatterplots of within gene estimates for one pair of replicates from
the simulated RNA-seq reads. The blue color corresponds to a smoothed color
density representation of the scatterplot and the red color emphasizes points
from the lowest regional densities.}
\label{fig:spanki2}
\end{figure*}

Figures \ref{fig:spanki1}
and \ref{fig:spanki2} illustrate the scatterplots of within gene estimates for
the first replicate versus the true values and the second replicate estimates,
respectively. The plots are ordered according to the inter-replicate consistency Mean Absolute Error, as shown in Figure \ref{fig:spanki2}. Note that RSEM-PME, BitSeqMCMC and BitSeqVB are the only methods which
avoid extreme outliers on the boundary of these graphs.


\subsection{Scenario 3: mixture of Poisson GLMs}

\begin{figure*}[t]
\centering
\includegraphics[width=0.9\textwidth]{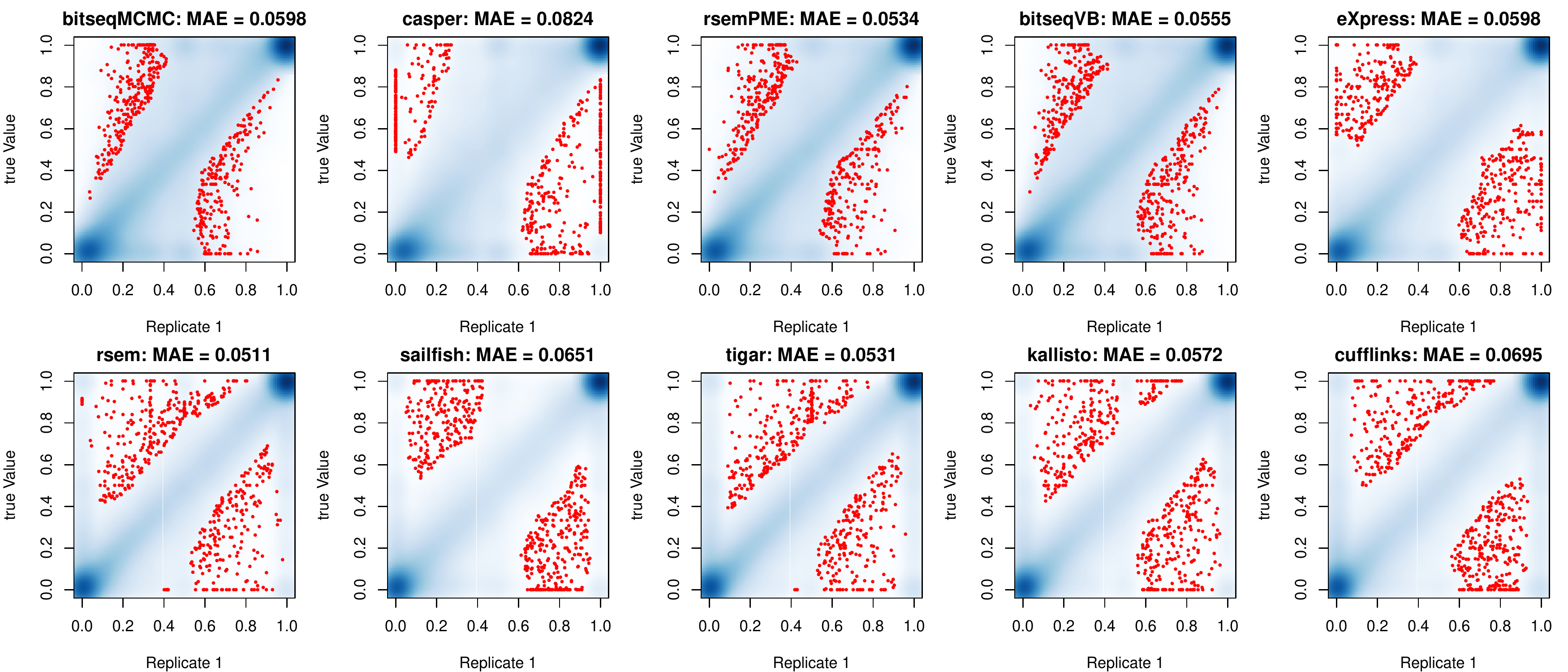}\\
\caption{Scenario 3: Scatterplots of within gene estimates for one replicate of the
simulated data versus the true values. The blue color corresponds to a smoothed
color density representation of the scatterplot and the red color emphasizes
points from the lowest regional densities.}
\label{fig:poismix-spanki1}
\end{figure*}
\begin{figure*}[ht]
\centering
\includegraphics[width=0.9\textwidth]{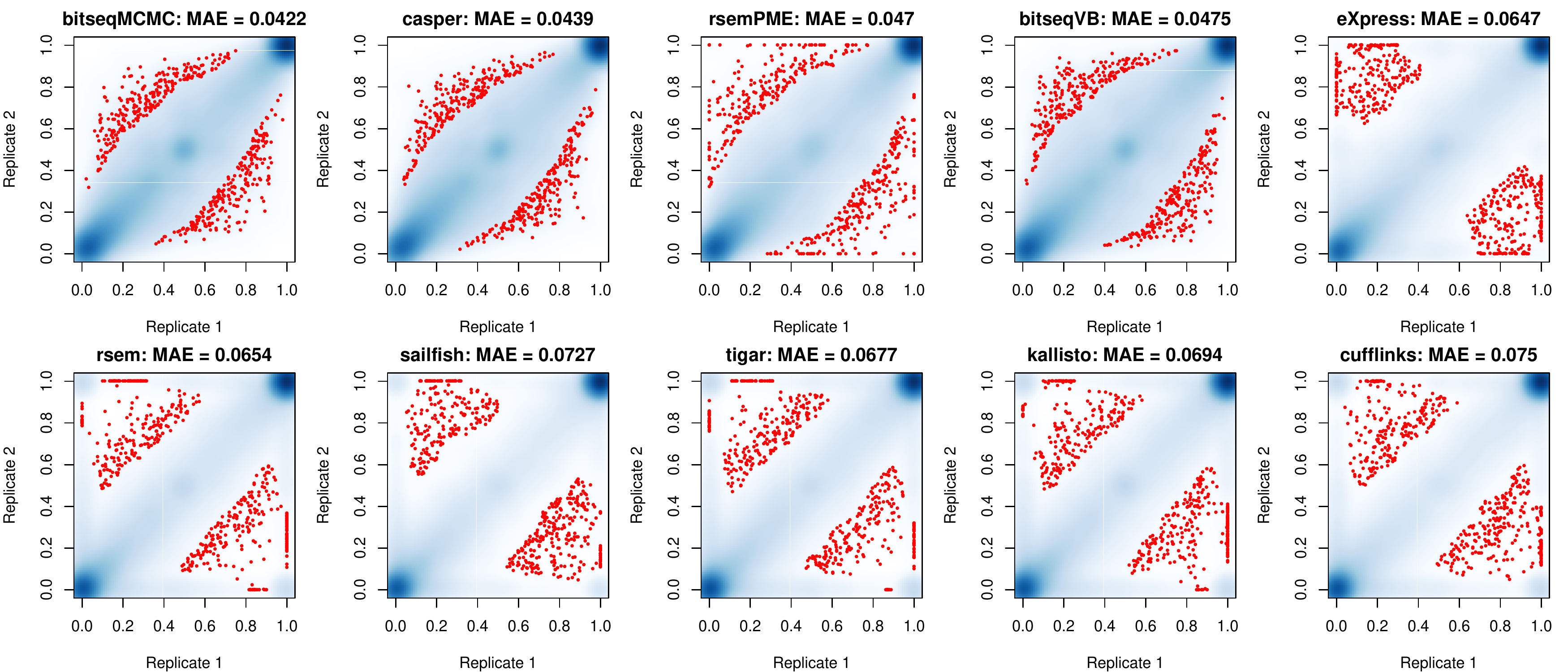}\\
\caption{Scenario 3: Scatterplots of within gene estimates for one pair of replicates from
the simulated RNA-seq reads. The blue color corresponds to a smoothed color
density representation of the scatterplot and the red color emphasizes points
from the lowest regional densities.}
\label{fig:poismix-spanki2}
\end{figure*}

A ground truth was generated using a mixture of Poisson Generalized Linear models. It has been recently demostrated that this approach can be used in order to model the underlying heterogeneity in RNA-seq datasets \citep{poisGLM}. Let $\mathcal P(\lambda)$ denotes the Poisson distribution with mean $\lambda > 0$ and let also $\mathcal D(\alpha_1,\ldots,\alpha_K)$ denotes the Dirichlet distribution with $\alpha_k>0$, $k = 1,\ldots,K$ for a given positive integer $K$. Reads were simulated according to the following generative process:
\begin{eqnarray*}
K &=& 20 \\
(\pi_1,\ldots,\pi_K) &\sim& \mathcal D(1,\ldots,1)\\
x_m &\sim& \mathcal U(0,30)\\
\log\lambda_{jkm} &=& \alpha_{k} + \beta_{jk} x_m\\
(\mu_{1m},\ldots,\mu_{5m}) &\sim& \sum_{k=1}^{K}\pi_k\prod_{j=1}^{5}\mathcal P(\lambda_{jkm}),\\
\mbox{RPK}_{jm} &\sim& \mathcal{NB}(\mu_{jm},50),
\end{eqnarray*}
where $m=1,\ldots,M$, $j = 1,\ldots,5$ and $k = 1,\ldots,K$. The regression coefficients ($\alpha_{k},\beta_{jk}$) are generated as follows.
\begin{eqnarray*}
\alpha_{k}&=&\frac{K^*-k}{2}, \quad k=1,\ldots,K, \\
\beta_{jk}&=& \begin{cases} \frac{10-\alpha_{k}}{10} , & k=1,\ldots,K,\: j=1 \\ \beta_{1k} + \varepsilon_{jk}, & j\geqslant 2, \: \varepsilon_{jk}\sim\mathcal N(0,0.001^2) \end{cases}
\end{eqnarray*} 
where $K^*:=\left\lfloor 0.5+(K+1)/2\right\rfloor$
($\left\lfloor \alpha\right\rfloor$ stands for the integer part of $\alpha$).

This resulted in $5.5$ million paired-end reads  of 76 base-pairs per
replicate, almost $27.5$ million reads in total. Figures \ref{fig:poismix-spanki1}
and \ref{fig:poismix-spanki2} illustrate the scatterplots of within gene estimates for
the first replicate versus the true values and the second replicate estimates,
respectively. The plots are ordered according to the inter-replicate consistency Mean Absolute Error, as shown in Figure \ref{fig:poismix-spanki2}. Note that BitSeqMCMC, BitSeqVB and Casper are the only methods which
avoid extreme outliers on the boundary of the inter-replicate consistency graphs.

\subsection{Scenario 4: RSEM estimates from real data}

\begin{figure*}[t]
\centering
\includegraphics[width=0.9\textwidth]{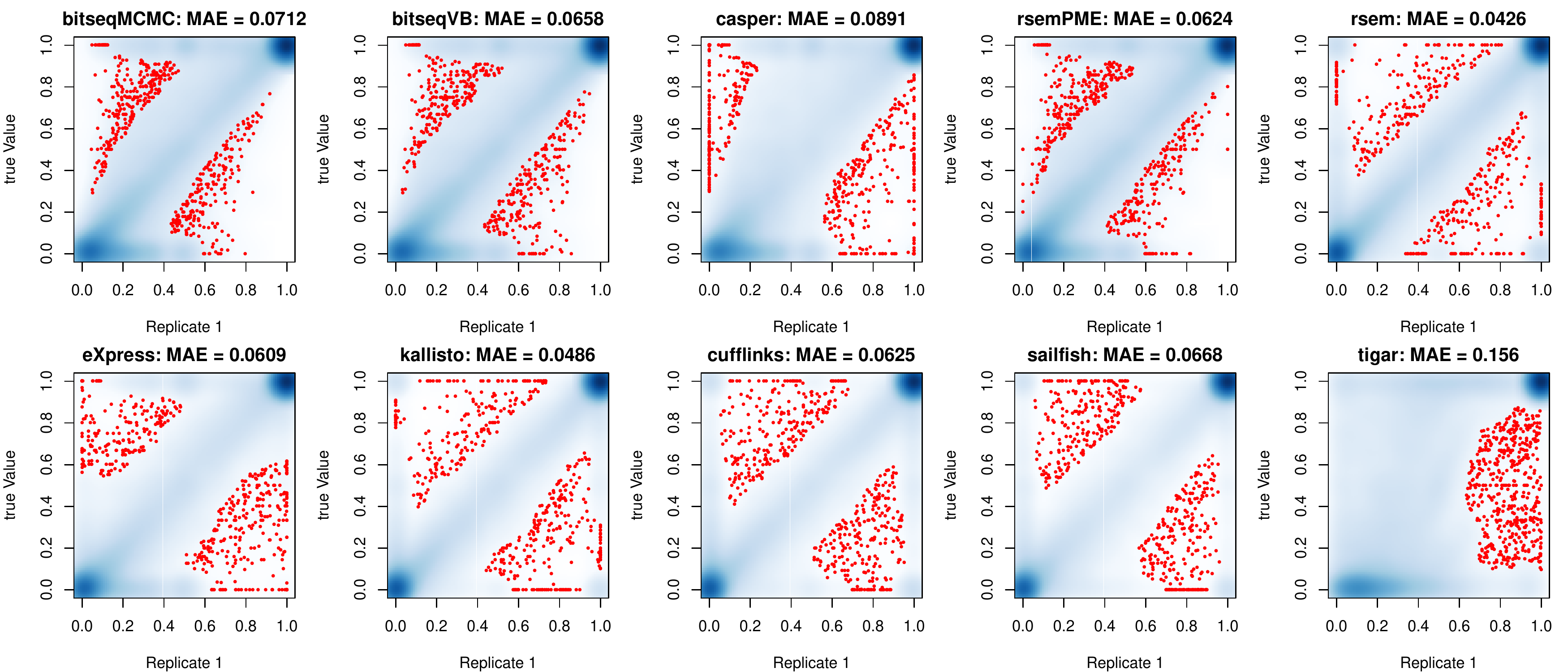}\\
\caption{Scenario 4: Scatterplots of within gene estimates for one replicate of the
simulated data versus the true values. The blue color corresponds to a smoothed
color density representation of the scatterplot and the red color emphasizes
points from the lowest regional densities.}
\label{fig:rsem-spanki1}
\end{figure*}
\begin{figure*}[ht]
\centering
\includegraphics[width=0.9\textwidth]{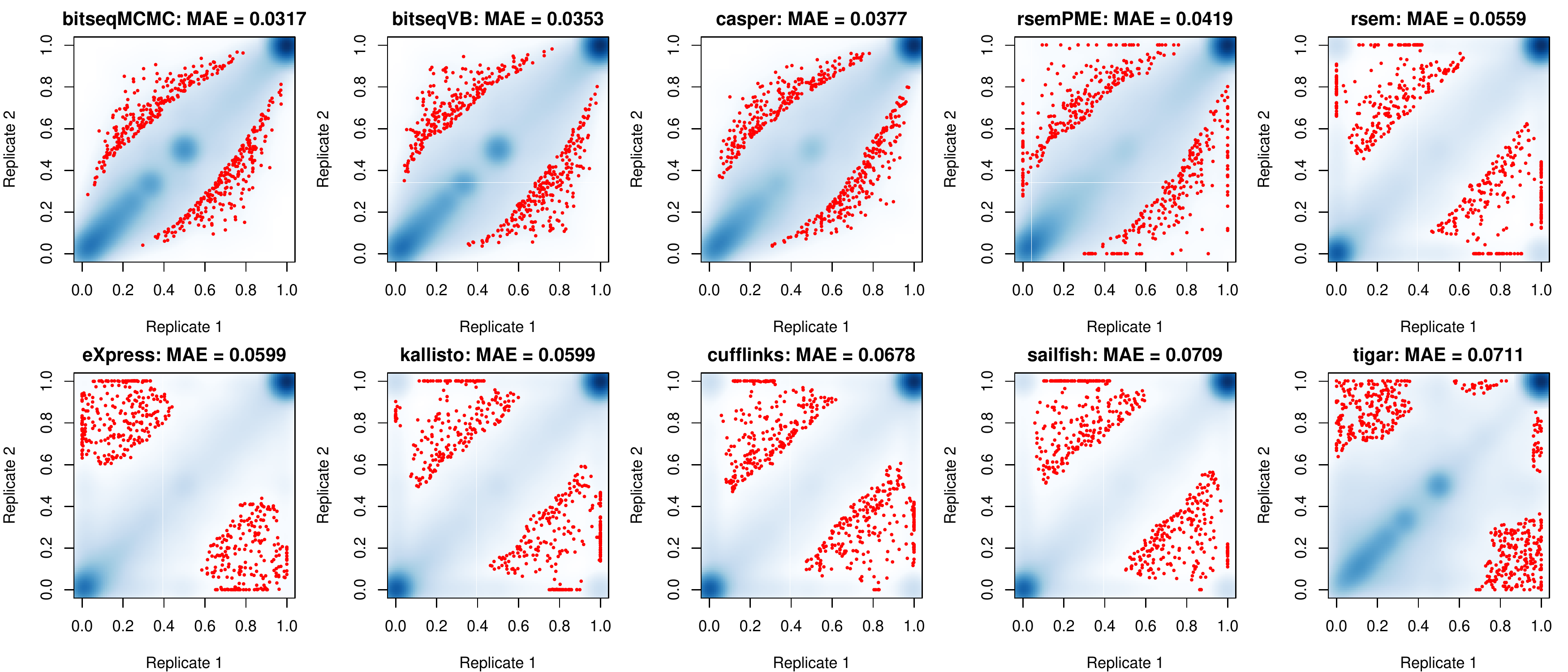}\\
\caption{Scenario 4: Scatterplots of within gene estimates for one pair of replicates from
the simulated RNA-seq reads. The blue color corresponds to a smoothed color
density representation of the scatterplot and the red color emphasizes points
from the lowest regional densities.}
\label{fig:rsem-spanki2}
\end{figure*}

We used human data (replicates SRR307907 and SRR307908) from the ENCODE project in order to capture the dynamics of realistic RNA-seq datasets. RSEM was used to estimate the relative expression levels of $M=48009$ transcripts for each dataset. Next, the resulting estimates were averaged and used as input to generate the baseline mean. Reads were simulated according to the following generative process: 
\begin{eqnarray*}
\mbox{RPK}_{jm} &\sim& \mathcal{NB}(\widehat\mu_{m},50),\quad m=1,\ldots,M, j = 1,\ldots,5.
\end{eqnarray*}
with $\widehat\mu_m$, $m=1,\ldots,M$ denoting the corresponding estimates of RPK values according to RSEM. This resulted in $18$ million paired-end reads  of 76 base-pairs per
replicate, almost $86$ million reads in total.  

Figures \ref{fig:rsem-spanki1}
and \ref{fig:rsem-spanki2} illustrate the scatterplots of within gene estimates for
the first replicate versus the true values and the second replicate estimates,
respectively. The plots are ordered according to the inter-replicate consistency Mean Absolute Error, as shown in Figure \ref{fig:rsem-spanki2}. Note that BitSeqMCMC, BitSeqVB and Casper are the only methods which
avoid extreme outliers on the boundary of the inter-replicate consistency graphs. 

\begin{figure*}[ht]
\centering
\begin{tabular}{c}
\includegraphics[width=0.95\textwidth]{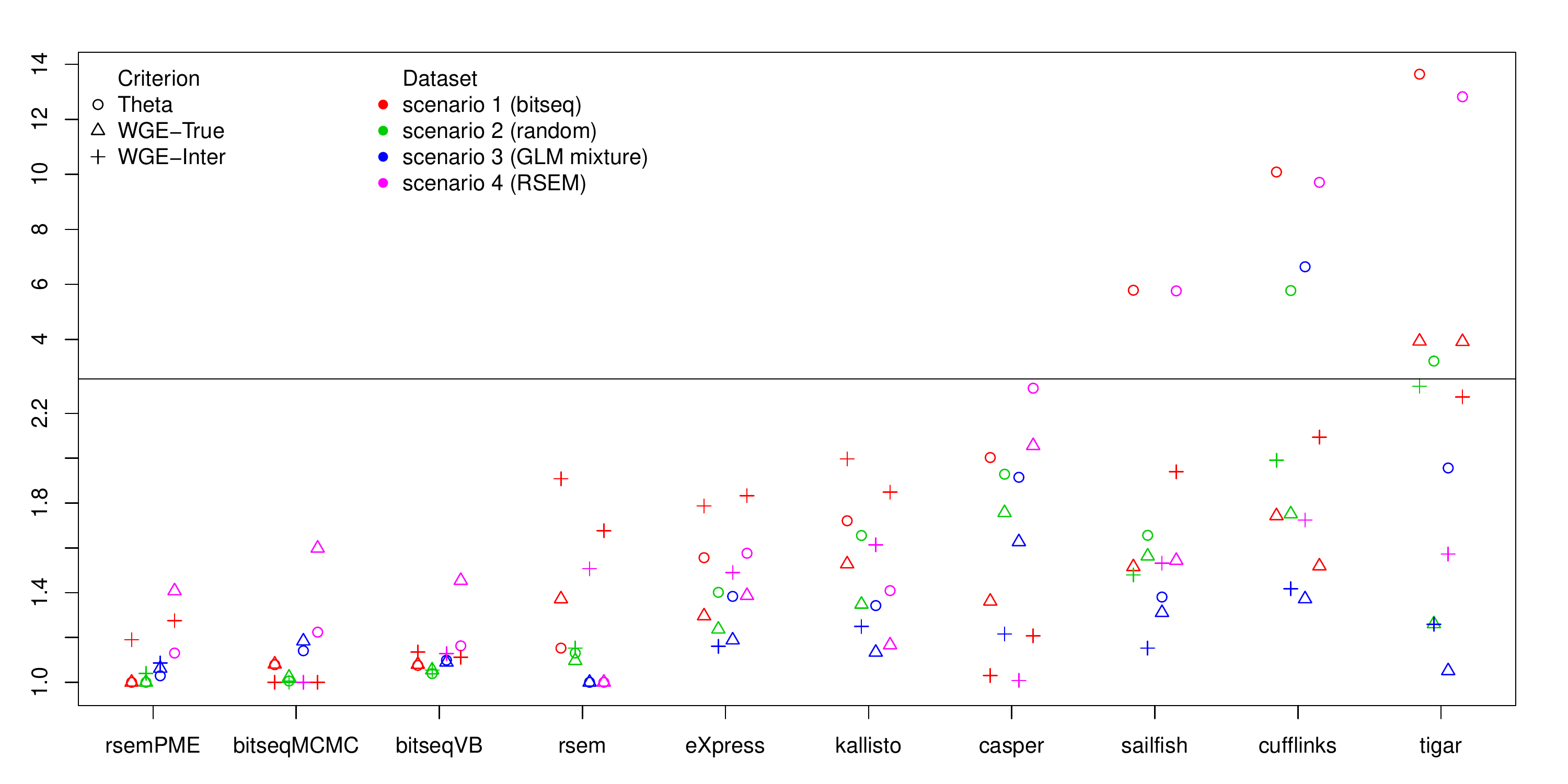}\\
\end{tabular}
\vspace{0.0em}
\caption{Ranking of methods for five replicates of simulated RNA-seq reads.
WGE-Inter: inter-replicate consistency of within gene estimates, WGE-True:
within gene estimates compared to the true values and Theta: estimated relative
transcript expression compared to the true values. Scores have been normalised by dividing by the min score per criterion.}
\label{fig:sim1-min}
\end{figure*}

The overall ranking of methods is shown in Figure \ref{fig:sim1-min}. Now we used an alternative normalisation by dividing each score by the minimum per criterion, so that the normalised score of the best method is equal to 1. 

\section{Evaluation measures}

The following measures are used.

\paragraph{Theta:}{$\frac{1}{J}\sum_{j=1}^{J}\frac{1}{M}\sum_{m=1}^{M}|\log\widehat{\theta}_m^{(j)} - \log\theta_m^{(j)}|$, where $\theta_m^{(j)}$ and $\widehat{\theta}_m^{(j)}$ denote true and estimated relative transcript expression for transcript $m$ of replicate $j$, $m = 1,\ldots,M$, $j=1,\ldots,J$.}
\paragraph{WGE-True:}{$\frac{1}{J}\sum_{j=1}^{J}\frac{1}{M}\sum_{m=1}^{M}|\widehat{q}_m^{(j)} - q_m^{(j)}|$, where $q_m^{(j)} := \frac{\theta_m^{(j)}}{\sum_{k\in T_m}\theta_m^{(j)}}$ and $\widehat{q}_m^{(j)} := \frac{\widehat{\theta}_m^{(j)}}{\sum_{k\in T_m}\widehat{\theta}_m^{(j)}}$ denote the true and estimated relative within gene transcript expression levels and also $T_m:=\{k:\mbox{ transcript \textit{k} in same gene as transcript \textit{m}, } k = 1,\ldots,M\}$, $m = 1,\ldots,M$ denotes the set of transcripts belonging to the parent gene of each transcript.}
\paragraph{WGE-Inter:}{$\sum_{i < j}\frac{2}{J(J-1)}\sum_{m=1}^{M}\frac{1}{M}\sum_{m=1}^{M}|\widehat{q}_m^{(i)} - \widehat{q}_m^{(j)}|$.}

\begin{figure*}[t]
\centering
\includegraphics[width=0.9\textwidth]{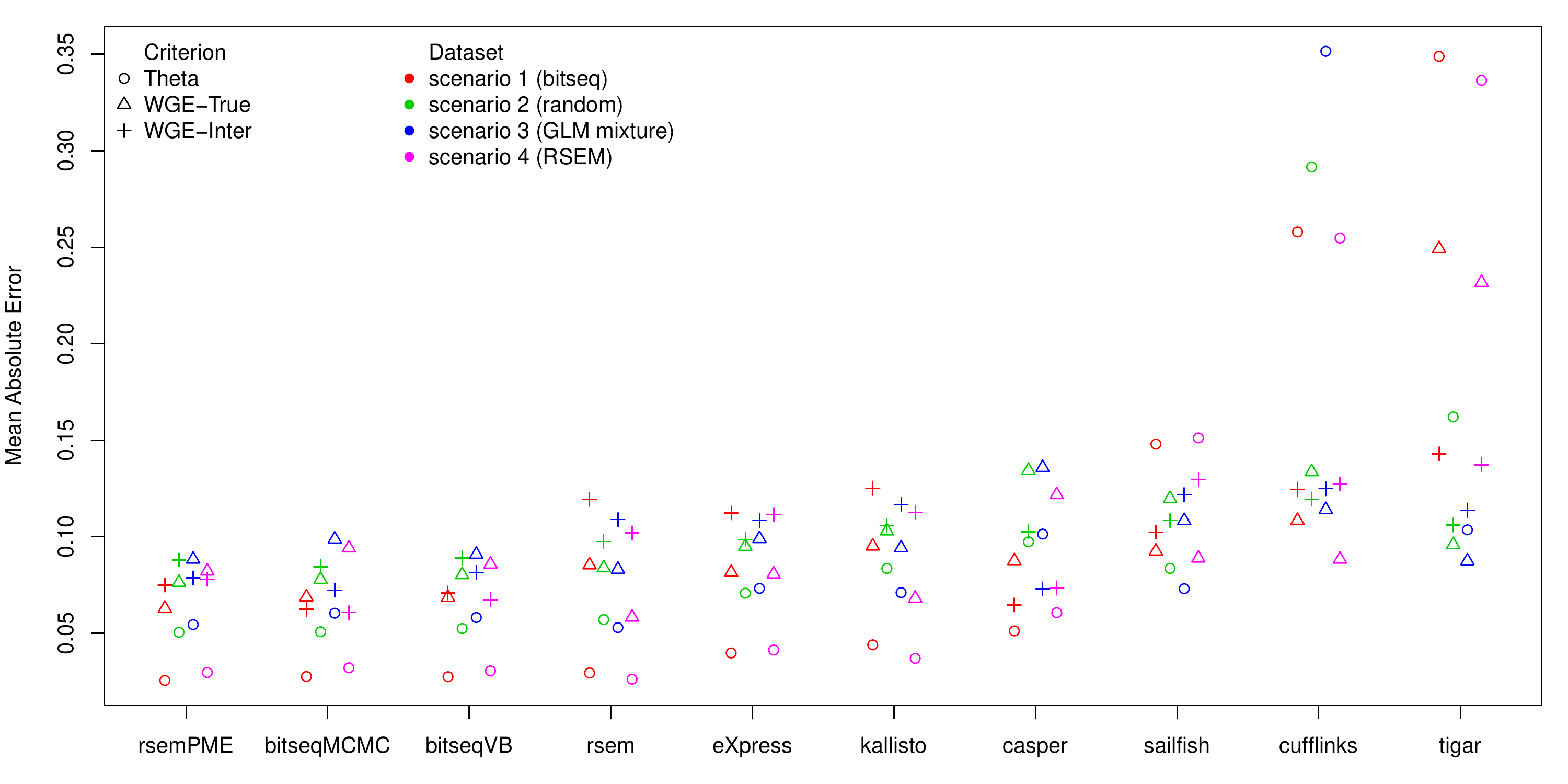}\\
\includegraphics[width=0.9\textwidth]{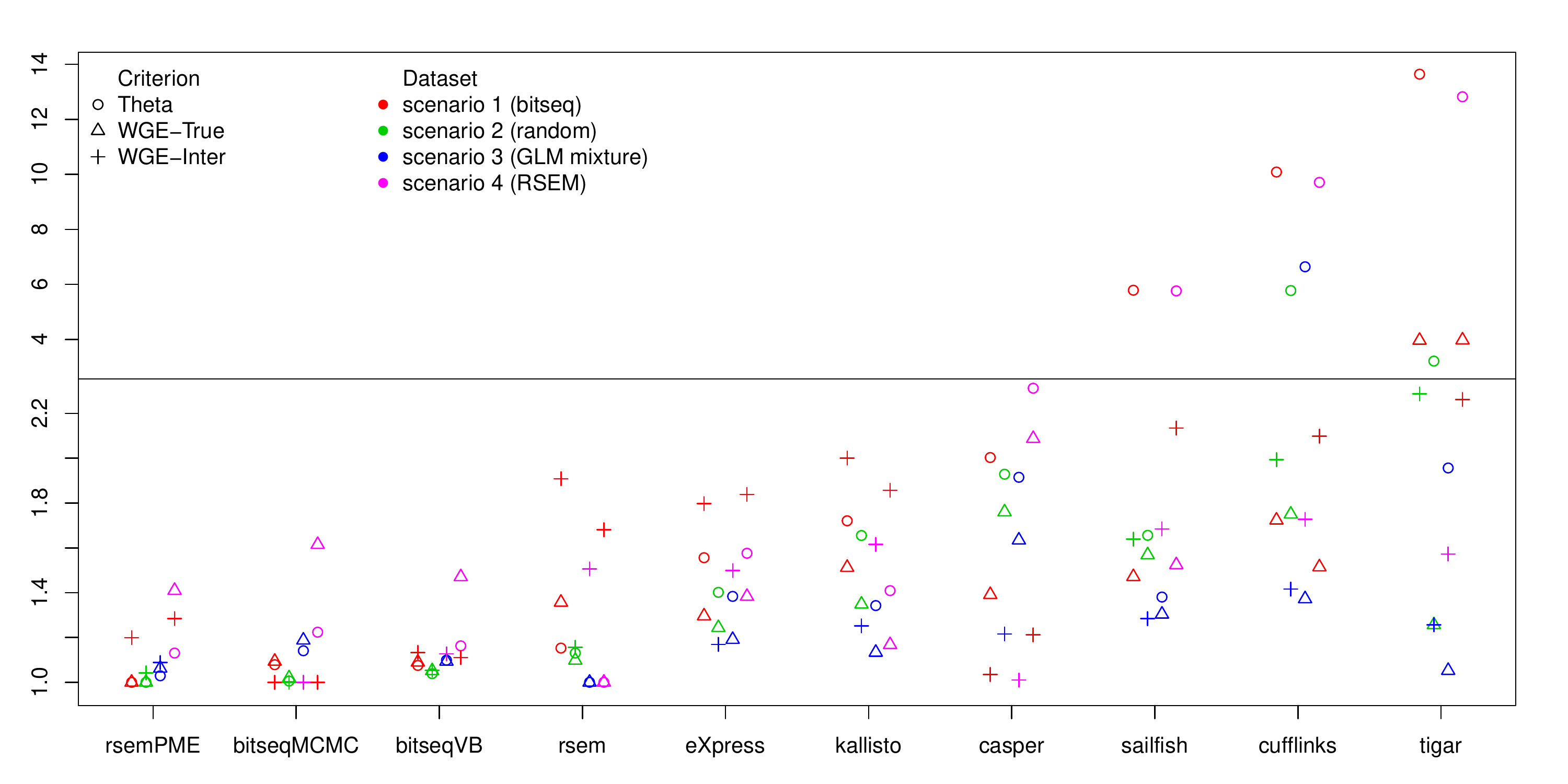}
\caption{Ranking of methods when using the TPM instead of theta for the within gene estimates (up: normalization according to unity sum, down: normalization by dividing by the min).}
\label{fig:TPM}
\end{figure*}

\begin{figure*}[t]
\centering
\includegraphics[width=0.9\textwidth]{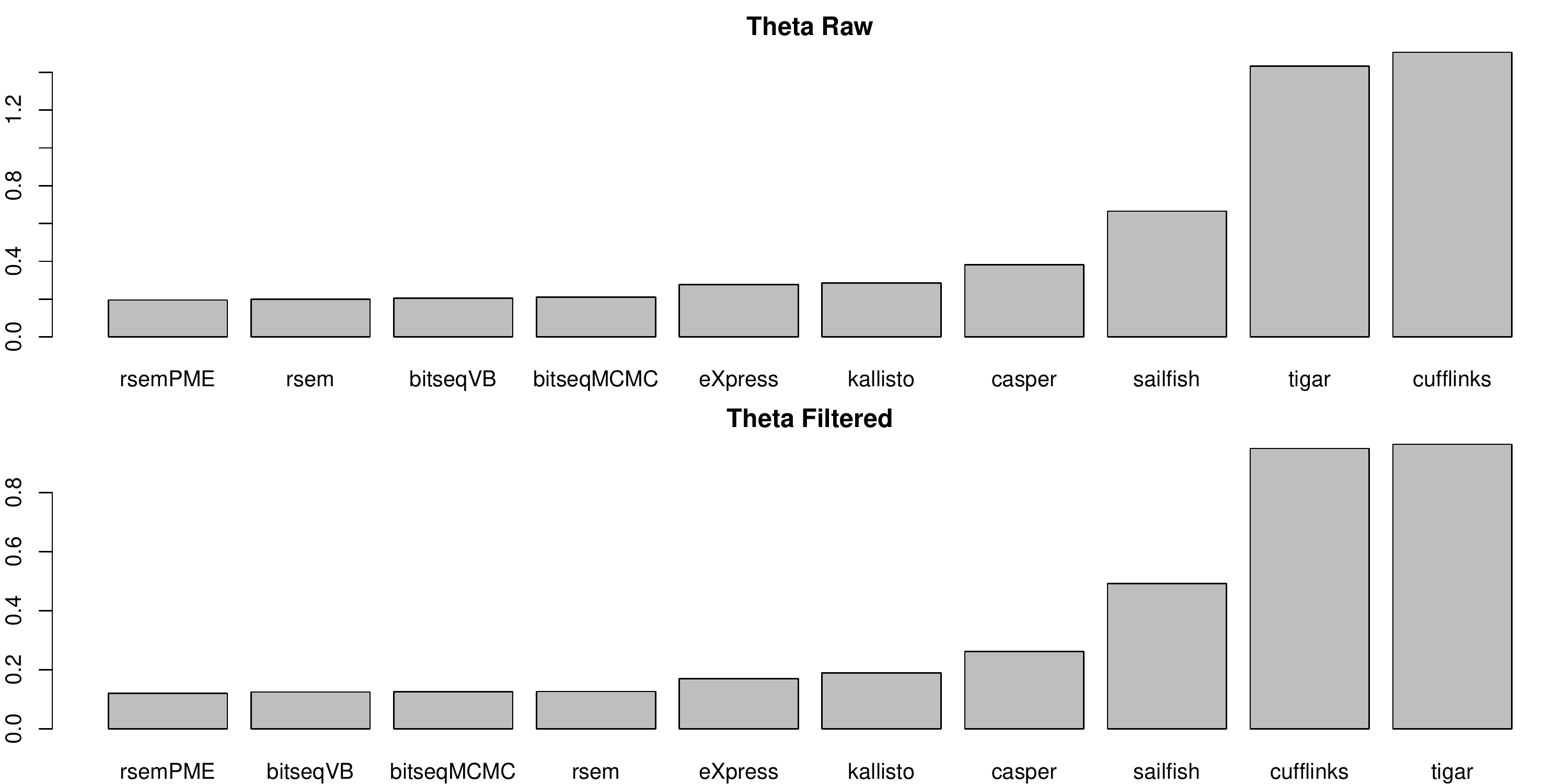}
\caption{Ranking of methods according to Theta when using all expressed transcripts (up) and only the ones with at least 50 reads per replicate (down).}
\label{fig:filter}
\end{figure*}

In addition we provide in Figure \ref{fig:TPM} the ranking of methods when the within gene estimates are computed according to the TPM measure, instead of $\theta$. Figure \ref{fig:filter} displays a comparison of ranking of methods when using only a subset of highly expressed transcripts. 

\begin{figure*}[ht]
\centering
\begin{tabular}{c}
\includegraphics[width=0.75\textwidth]{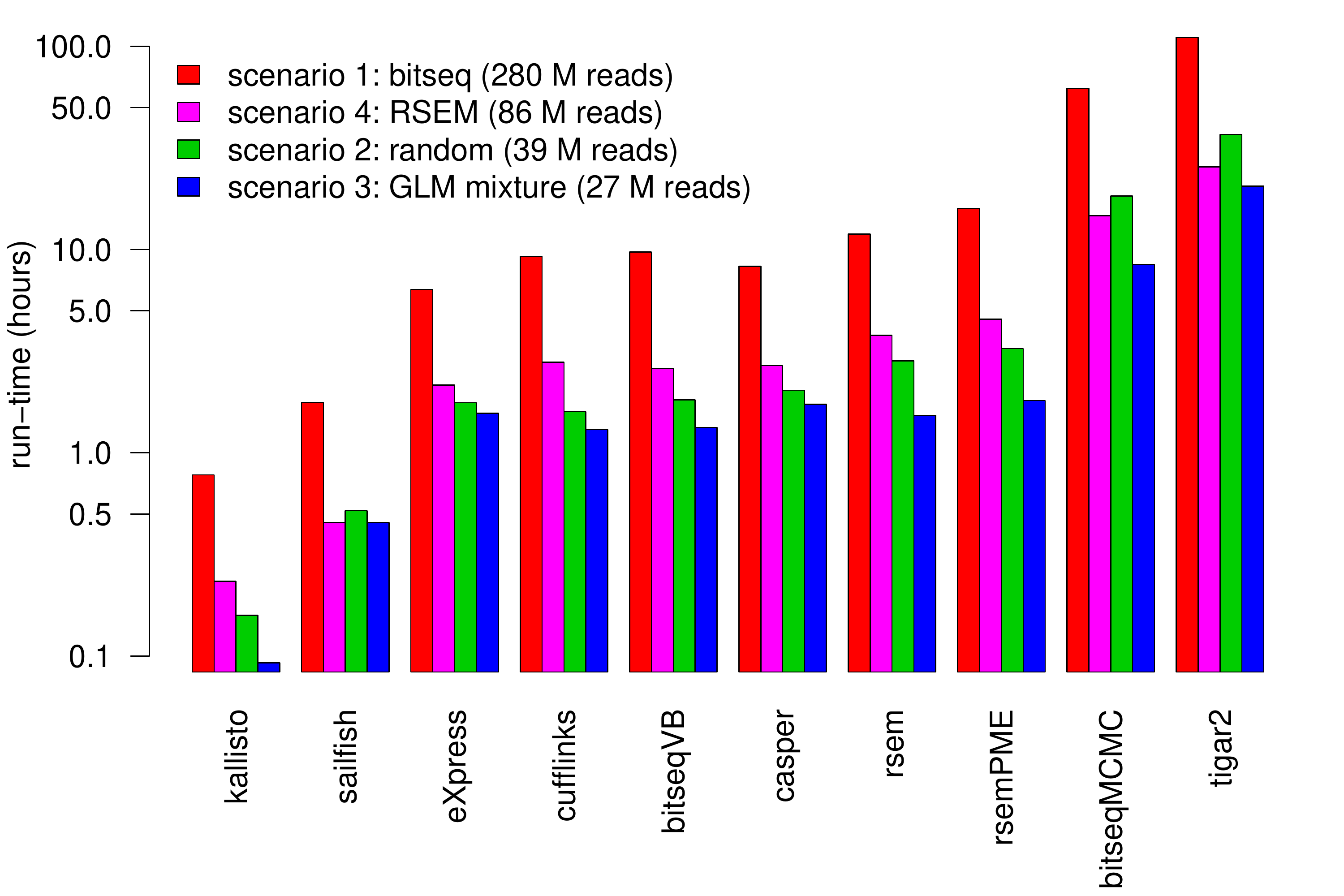}\\
\end{tabular}
\vspace{0.0em}
\caption{Runtime per method including the time needed for the alignment procedure.}
\label{fig:sim1-mapTimes}
\end{figure*}

\bibliographystyle{apalike} 
\bibliography{references}      

\end{document}